\documentclass[%
 aip,jcp,
 amsmath,amssymb,
preprint,
]{revtex4-2}

\usepackage{graphicx}
\usepackage{caption}
\usepackage{subcaption}
\usepackage{pdflscape}
\usepackage{afterpage}
\usepackage{capt-of}

\usepackage{floatpag}

\usepackage{dcolumn}
\usepackage{bm}

\usepackage[utf8]{inputenc}
\usepackage[T1]{fontenc}
\usepackage{booktabs}

\usepackage{mhchem}

\makeatletter
\AtBeginDocument{\let\LS@rot\@undefined}
\makeatother

\usepackage{todonotes}
\usepackage{braket}

\DeclareUnicodeCharacter{2212}{-}
\DeclareUnicodeCharacter{2009}{ }
\begin{document}

\title{Explicitly correlated coupled cluster method for accurate treatment of
open-shell molecules with hundreds of atoms}

\author{Ashutosh Kumar} 
\affiliation{Department of Chemistry, Virginia Tech, Blacksburg, Virginia 24061, United States}

\author{Frank Neese}
\affiliation{Max-Planck-Institut f{\"u}r Kohlenforschung, Kaiser-Wilhelm-Platz 1, D-45470 M{\"u}lheim an der Ruhr, Germany}

\author{Edward F. Valeev}
 \email{efv@vt.edu}
\affiliation{Department of Chemistry, Virginia Tech, Blacksburg, Virginia 24061, United States}

\date{\today} 

\begin{abstract}
We present a near-linear scaling formulation of the explicitly-correlated coupled-cluster singles and doubles with perturbative triples method 
(CCSD(T)$_{\overline{\text{F12}}}$) for high-spin states of open-shell species. The approach is based on the conventional open-shell CCSD formalism 
[M. Saitow et al., J. Chem. Phys. 146, 164105 (2017)] utilizing the domain local pair-natural orbitals (DLPNO) framework. The use of spin-independent 
set of pair-natural orbitals ensures exact agreement with the closed-shell formalism reported previously, with only marginally impact on the cost 
(e.g. the open-shell formalism is only 1.5 times slower than the closed-shell counterpart for the \ce{C160H322} n-alkane, with the measured size 
complexity of $\approx1.2$). Evaluation of coupled-cluster energies near the complete-basis-set (CBS) limit for open-shell systems with more than 
550 atoms and 5000 basis functions is feasible on a single multi-core computer in less than 3 days. The aug-cc-pVTZ DLPNO-CCSD(T)$_{\overline{\text{F12}}}$ 
contribution to the heat of formation for the 50 largest molecules among the 348 core combustion species benchmark set [J. Klippenstein et al., J. Phys. Chem. A 121, 6580 (2017)] had root-mean-square deviation (RMSD) from the extrapolated CBS CCSD(T) reference values of 0.3 kcal/mol. For a more challenging set of 50 reactions involving small closed- and open-shell molecules [G. Knizia et al., J. Chem. Phys. 130, 054104 (2009)] the aug-cc-pVQ(+d)Z DLPNO-CCSD(T)$_{\overline{\text{F12}}}$ yielded a RMSD of $\sim$0.4 kcal/mol with respect to the CBS CCSD(T) estimate.

\end{abstract}

\maketitle
\section{Introduction}
One of the central problems limiting the application of accurate {\em am initio} methods to large molecular systems is their high computational complexity.
For example, the operation and storage requirements of the coupled cluster singles and doubles with perturbative triples\cite{Raghavachari:1989} method (CCSD(T))
 --- considered to be the ``gold standard'' of quantum chemistry --- scale as ${\cal O}(N^7)$ and ${\cal O}(N^4)$ respectively, with $N$
 proportional to the system size (e.g. number of electrons or atoms).
However, this steep scaling is clearly unphysical, especially for insulators, since the dynamical electron correlation is a local phenomenon
\cite{Sinanoglu:1964,Nesbet:1965}; correlation energy of two well-separated electrons
decays as $R^{-6}$,  where $R$ is the distance between the centroids of the respective 1-electron states.
This problem arises due to the unstructured (thus, often delocalized) nature of the reference orbitals (e.g., Hartree-Fock (HF))  
which results in the appearance of many non-negligible wave function parameters.

The use of {\em localized}
occupied and  unoccupied orbitals allows to reveal the sparsity of the wave function and the Hamiltonian operator.
Whereas most localization methods for occupied orbitals result in similar performance, the established strategies for constructing
localized unoccupied orbitals differ strongly in the performance and technical complexity;
here we only focus on the two pertinent examples.
The projected atomic orbitals\cite{Pulay:1983,Saebo:1986,PulaySaebo:1993} (PAOs) introduced by Pulay and Saeb\o{}
are inexpensive to construct but due to their nonorthogonality and linear dependence require orthogonalization within each
orbital and pair ``domain''; the PAO technology was developed by Sch{\"u}tz, Werner and co-workers into a large toolkit of
linear scaling correlated methods\cite{Schutz:1999,Schutz:2000a,Schutz:2000bx,Schutz:2001,Schutz:2002dc}. Unfortunately robust application
of PAO-based methods requires the use of relatively large domains\cite{Riplinger:2013tn,Pinski:2015ii,Werner:2015kw,McAlexander16:LRCC} (20-30 atoms / electron pair) which pushes the crossover
with conventional methods to very large molecules.
Pair-natural orbitals (PNOs) is another representation of the unoccupied space  that for each electron pair
is far more compact than the comparable PAO pair domain due to the use of the approximate pair
correlation amplitudes. Originally introduced by Edmiston 
and Krauss in 1965\cite{Edmiston66}, the PNOs were developed in 1970s
by Meyer,\cite{Meyer73,Meyer74} Ahlrichs, Kutzelnigg, and Staemmler\cite{Ahlrichs75}.

Although regarded as impractical due to the costs of transforming the Hamiltonian to the PNO basis
the PNO method was revived by Neese and co-workers by leveraging the local
density-fitting and other approximations\cite{Neese:2009,Neese:2009db, 
Riplinger:2013tn, Riplinger:2013ek} also used to power the PAO-based methods.
The initial \textit{Local} PNO (LPNO) CC theory developed by some of us, after successive rounds of improvement has now evolved
into the state-of-the-art near-linear-scaling {\em Domain} based LPNO (DLPNO) CC\cite{Pinski:2015ii,Riplinger:2016} methods which are
characterized by the use of  (1) both PAO (intermediate) 
and PNO representations,  (2) Hilbert-space measures for distance between orbitals\cite{Pinski:2015ii},
(3) \textit{SparseMap}\cite{Pinski:2015ii} representation of
(block-)sparse tensors in terms of binary index-to-domain maps for compact specification of structure and algorithms,
(4) hierarchy of pair-screening approximations, etc; similar traits apply to the work of Werner {\em et al.} \cite{Werner:2015kw,Ma:2015gs} and Tew, Helmich and 
H{\"a}ttig\cite{Tew:2011,Hattig:2012}. These formalisms are fundamental and permit evaluation of other properties such as
analytic nuclear gradients\cite{Hattig:2017PNOMP2Grad,Neese:2019DLPNOMP2Grad}, excitation energies
\cite{Helmich:2013bd,Helmich:2014by,Hattig:2018PNOExE,KumarDutta2017,kallay:2019PNOADC,Dittmer2019,Salla:2019}, electron attachment/detachment energies\cite{Izsak:2018PNOION,
Izsak:2019DLPNOEA,Dutta:2019IPEOMDLPNO} and response properties\cite{McAlexander16:LRCC,CrawfordCCResponse:2019}.
It is also possible to apply the PNO formalism in the context of multi-configuration methods\cite{Neese:2015MkPNO,Werner:2016PNOCASPT2,Guo:2016DLPNONEVPT2,
Ondrej:2018MkDLPNO}.
The closed-shell DLPNO-CCSD(T) method can now regularly calculate 
energies of protein fragments consisting of a few thousand atoms on a single multi-core computer in a matter of few days\cite{Riplinger:2016}.
However, most of the interesting and useful chemical phenomena almost always involve open-shell molecular systems like organic radicals, 
transition metal compounds etc. For the sake of brevity it is not possible to recap the rich history of developments of the single-reference coupled-cluster methods
applicable to open-shell systems, without and with spin adaptation;\cite{Purvis:1982,Rittby:1988,Watts:1990,Janssen:1991ct,Watts:1993,Li:1994cv} interested readers are referred to a number of excellent reviews\cite{Musial:2007,Krylov:2008bh,Bartlett:2010,Bartlett:2012}
However, extension of the DLPNO-CC methods to the treatment of high-spin open-shell systems has been relatively recent. Hansen, Neese and co-workers extended the LPNO-CC methodology to 
the open-shell regime in 2011 using quasi-restricted orbitals (QROs)\cite{Neese:2006} as reference orbitals, with multiple sets of PNOs (for different spin components)\cite{Hansen:2011}, 
resulting in a number of technical problems like multiple integral transformations and PNO overlap matrices, unbalanced treatment of closed- and open-shell species etc. Saitow and 
co-workers were able to resolve a majority of these issues through the recently developed open-shell DLPNO-CCSD method\cite{Saitow:2017bo} by using a restricted 
open-shell HF (ROHF) framework, where the PNOs are generated through the diagonalization of pair-specific 1-electron reduced-density matrix (RDM) obtained from the 
second-order N-electron valence perturbation theory (NEVPT2). The advantages associated with this approach would be discussed specifically, later in the manuscript. Guo and 
co-workers recently extended the open-shell DLPNO-CCSD method to perturbative triples in a linear scaling fashion\cite{Yang:2020}. In similar works, Werner
implemented a parallel and efficient near-linear scaling open-shell spin-restricted MP2\cite{Werner:2019} and more recently, both (spin-) restricted and unrestricted
variants of CCSD model\cite{Werner:2020} where the PNOs were obtained through 1-RDMs of second-order M{\o}ller-Plesset perturbation theory (MP2).

However, reduction of the computational complexity
is not solely sufficient for practical application of reduced-scaling CC methods.
The slow convergence of molecular properties with respect to the basis set quality must also be addressed.
For example, the basis set error of atomic correlation energy
decreases as $\mathcal{O}[(L_\text{max}+1)^{-3}]$, for a basis set saturated to angular 
momentum $L_{\text{max}}$ due to the use of Slater determinants as the basis.\cite{Kutzelnigg:1992kj}
Explicitly correlated R12/F12 methods,\cite{Kutzelnigg1985R12,Klopper:2002vn,Manby:2003db,TenNo:2004,TenNo:2004dg,Valeev:2004ep,Kedzuch:2005hr,Fliegl:2005wb,Werner:2007ub,Valeev:2008,Zhang:2012it} 
in which explicit dependence on the interelectronic distances ($r_{ij}$) are included in the wave function, are a robust way to reduce the basis set error
with only a modest increase in the computational expenses, for example the MP2-R12 methods pioneered by Kutzelnigg\cite{Kutzelnigg1985R12} reduced the basis set errors in atoms to 
$\mathcal{O}[(L_\text{max}+1)^{-7}]$\footnote{However, further numerical approximations like the resolution of identity in the F12 methods can sometimes 
slow down this convergence. Furthermore, precise analysis of the basis set convergence of correlation energy for molecules is more phenomenological than for atoms.}.
This makes these methods an attractive alternative 
to the extrapolation based methods, which often require very large basis set calculations\cite{Liakos:2013}.

Unsurprisingly, recent years have seen numerous developments in the reduced-scaling explicitly correlated coupled-cluster methods.
Werner and Krause did a comparative analysis on different 
choices of representations of the unoccupied space (PAOs, PNOs, orbital specific virtuals (OSVs)\cite{Yang:2012kq}) in the explicitly correlated CCSD theory\cite{Krause:2012fx}. 
Tew and H{\"a}ttig implemented the reduced-scaling variants of explicitly correlated MP2 and CCSD methods utilizing PNOs\cite{Tew:2011,Hattig:2012,Tew:2013,Schmitz:2014}. 
Werner and co-workers recently came up with an efficient parallel implementation of explicitly correlated PNO based MP2 and CCSD methods\cite{Ma:2015gs,Ma:2018}. 
Some of us recently implemented the near-linear-scaling DLPNO-MP2-F12\cite{Pavosevic:2016bc} and DLPNO-CCSD(T)$_{\overline{\text{F12}}}$ methods\cite{Pavosevic:2017kb} utilizing the \textit{SparseMap} infrastructure.

Unfortunately, the PNO-based explicitly correlated coupled-cluster methods to open-shell species are not yet readily available.
Werner and co-workers recently used a simulation code to test the efficiency of the PNOs obtained from his PNO-RMP2 model in correlated PNO-RCCSD-F12 calculations\cite{Werner:2019}.
Here we report the first {\em production-quality} implementation
of the DLPNO-CCSD(T)$_{\overline{\text{F12}}}$ method for open-shell high-spin molecular species.
It leverages the existing open-shell DLPNO-CCSD(T) method and 
the closed-shell DLPNO-F12 infrastructure while maintaining the near-linear scaling behavior observed for the closed-shell calculations.
This manuscript is structured as follows: Section \ref{sec:methods} introduces the formalism of explicitly correlated DLPNO-CCSD(T). 
Computational details are provided in Sec. \ref{sec:CompDetails}. In Sec. \ref{sec:Results}, we assess the performance of the new method with respect to robustness, scaling with the system size, and precision.
We summarize our findings in Sec. \ref{sec:Summary}.

\section{Formalism}\label{sec:methods}
This work utilizes the DLPNO framework to formulate a near-linear scaling perturbative coupled-cluster F12 model, 
CCSD(T)$_{\overline{\text{F12}}}$\cite{Zhang:2012it,CrawfordValeev:2008}.
Some of us implemented the closed-shell DLPNO-CCSD(T)$_{\overline{\text{F12}}}$ method in the ORCA quantum chemistry package\cite{Neese:2011ki} in 2018\cite{Pavosevic:2017kb} and here we extend it
to the case of high-spin open-shell states. This work can also be seen as a reduced-scaling 
extension of the open-shell canonical CCSD(T)$_{\overline{\text{F12}}}$ method, implemented by Zhang and Valeev\cite{Zhang:2012it}. 
In this section, we take a brief look at the formalism of CCSD(T)$_{\overline{\text{F12}}}$ 
along with the DLPNO approximations.

\subsection{The PNO-based CCSD(T)$_{\overline{\text{F12}}}$ Method}
The CCSD(T)$_{\overline{\text{F12}}}$ energy consists of four contributions:
\begin{equation}
E_{{\text{CCSD(T)}}_{\overline{\text{F12}}}} = E_{\text{CCSD}} + E_{\text{(T)}} + E_{{{\text{(2)}}_\text{S}}} 
+ E_{\text{{(2)}}_{\overline{\text{F12}}}},
\label{eq:energy}
\end{equation}
where the first two terms sum up to the conventional CCSD(T) energy, and the third and fourth terms 
correct for the basis set incompleteness of the reference and CCSD correlation energies, respectively.
The CCSD(T) correlation energy for open-shell species has already been implemented recently by 
Guo et al.\cite{Yang:2020} in a near-linear scaling fashion;
since the $E_{{{\text{(2)}}_\text{S}}}$ does not require reduced-scaling approximations,
we thus focus solely on the reduced-scaling formulation of 
the last term, $E_{{\text{(2)}}_{\overline{\text{F12}}}}$.

The $E_{{\text{(2)}}_{\overline{\text{F12}}}}$ energy is 
obtained via the second-order Hylleraas functional\cite{Kong:2012dx}:
\begin{equation}
E_{{\text{(2)}}_{\overline{\text{F12}}}} = \langle1_{\text{F12}}|{\hat{H}}^{(0)}|1_{\text{F12}}\rangle 
+ \langle\bar{0}|{\hat{H}}^{(1)}|1_{\text{F12}}\rangle \\
+ \langle1_{\text{F12}}|{\hat{H}}^{(1)}|\bar{0}\rangle,
\label{eq:Hylleraas}
\end{equation}
where $|\bar{0}\rangle$ and $\langle\bar{0}|$ are the right and left CCSD wave functions respectively and 
$|0\rangle$ refers to the reference determinant,
\begin{equation}
|\bar{0}\rangle = \text{exp}(\hat{T})|0\rangle
\label{eq:CCRwfn}
\end{equation}
\begin{equation}
\langle\bar{0}| = \langle0|(1+\hat{\Lambda})\text{exp}(-\hat{T})|.
\label{eq:CCLwfn}
\end{equation}
Using the notation of $i,j,k,..$ for active occupied orbitals and $a,b,c,..$ for unoccupied orbitals, 
the $\hat{T}$ operator within the framework of DLPNO-CCSD theory can be written as,
\begin{equation}
\hat{T} \equiv \hat{T}_1 + \hat{T}_2 \equiv t^i_{a_i} \tilde{a}^{\thinspace a_i}_{i} 
+ \frac{1}{{(2!)}^2} \thinspace t^{ij}_{a_{ij}b_{ij}} \tilde{a}^{\thinspace a_{ij}b_{ij}}_{ij}.
\label{eq:TOperator}
\end{equation}
where $\tilde{a}^{\thinspace a_i}_{i}$ and $\tilde{a}^{\thinspace a_{ij}b_{ij}}_{ij}$ are the one 
and two-particle excitation operators, $a_{ij}$ refer to the (doubles) PNOs
associated with the orbital pair $ij$, and ${a_i}$ refer to the orbital-specific unoccupied orbitals obtained as the PNOs of the ``diagonal'' $ii$ pairs (these
``singles PNOs'' are screened tighter than the standard doubles PNOs).
As in the conventional CCSD(T) method itself, the $\hat{\Lambda}$ operator (which makes the CCSD Lagrangian 
variational w.r.t. wave function parameters\cite{Helgaker:2000db}) is approximated by $\hat{T}^\dagger$.\cite{Stanton:1997vq}
However, it should be noted that the $\Lambda$-CCSD(T) method where one needs to solve for an additional set of $\hat{\Lambda}$ amplitude
equations, gives more accurate energies than the conventional CCSD(T) method, specially when the molecules are far from their equillibrium 
geometries\cite{Taube:2008}, albeit at almost twice the computational cost.
In the CCSD(T)$_{\overline{\text{F12}}}$ method, a L{\"o}wdin partitioning of the similarity transformed 
CCSD Hamiltonian matrix $\bar{H} \equiv \text{exp}(-\hat{T})\hat{H}\text{exp}(\hat{T})$ is carried out 
where the zeroth-order Hamiltonian in Eq.~(\ref{eq:Hylleraas}) is chosen to be the Fock operator $(\hat{F})$ while $\hat{H}^{(1)}$ is 
defined as $\bar{H}$. This choice ensures no coupling between the $E_{\text{(T)}}$ and $E_{\text{{(2)}}_{\overline{\text{F12}}}}$ 
corrections\cite{CrawfordValeev:2008} and thus the CCSD(T) energy is not affected by the presence of explicitly-correlated terms.
This makes the CCSD(T)$_{\overline{\text{F12}}}$ formalism quite favorable formally and computationally compared to the iterative CC-R12 methods
\cite{Fliegl:2005wb,Fliegl:2006,Werner:2007,Tew:2008,Shiozaki:2008,Hattig:2010gf}.
The perturbative F12 models also offer advantages for incorporating orbital optimizations (see 
the recent work of Kats and Tew\cite{kats:2019}). 
However, for systems where the coupling between the $\hat{T}$ amplitudes and the F12 geminal operator is strong the iterative CC-R12 methods 
offer a clear advantage, as the perturbative F12 models treat such coupling to finite, rather than infinite, order. The first-order explicitly 
correlated wave function (Eq.~(\ref{eq:Hylleraas})) can be expressed in terms of geminal functions,
\begin{equation}
|1_{\text{F12}}\rangle \equiv \frac{1}{2!} \sum_{ij} |\hat{\gamma}^{\thinspace ij}_{ij}\rangle, 
\label{eq:ExpCorrWfn}
\end{equation}
\begin{equation}
|\hat{\gamma}^{\thinspace ij}_{ij}\rangle = \frac{1}{2!}\bar{R}^{\thinspace ij}_{\alpha\beta} 
\tilde{a}^{\thinspace \alpha\beta}_{ij}|0\rangle,
\label{eq:ExpCorrWfnRbar}
\end{equation}
where $\bar{R}^{\thinspace ij}_{\alpha\beta}$ refers to the antisymmetrized form of the matrix element 
of the geminal correlation factor $f(r_{12})$:
\begin{equation}
R^{ij}_{\alpha\beta} \equiv \langle\alpha\beta|\hat{Q}_{ij}f(r_{12})|ij\rangle.
\label{eq:ExpCorrWfnR}
\end{equation}
The indices $\alpha,\beta$ correspond to the full unoccupied space, thus, the first-order
wave function of Eq.~(\ref{eq:ExpCorrWfn}) only consists of pure double excitations and 
is orthogonal to the reference wave function. Furthermore, the projection operator $\hat{Q}_{ij}$
makes the overlap of the wave function with the standard CCSD excitations to be zero,
\begin{equation}
\hat{Q}_{ij} = 1 - \sum_{a_{ij}b_{ij}} |a_{ij}b_{ij}\rangle\langle a_{ij}b_{ij}|.
\label{eq:Qij}
\end{equation}
An alternative projector based on PNO-like geminal spanning orbitals (GSOs) developed 
by some of us previously is not used in this work\cite{Pavosevic:2014}.
In order to satisfy the first-order cusp condition for the singlet (S=0) and triplet (S=1) 
pair functions, spin projectors for both the states are added to the geminal correlation factor 
along with the corresponding geminal amplitudes of $\frac{1}{2}$ (singlet) and $\frac{1}{4}$ (triplet):
\begin{equation}
R^{ij}_{\alpha\beta} \equiv \langle\alpha\beta|\hat{Q}_{ij}f(r_{12})(\frac{1}{2}\hat{P}_0 + \frac{1}{4}\hat{P_1})|ij\rangle.
\label{eq:ExpCorrWfnRspin}
\end{equation}
The projectors $\hat{P}_0$ and $\hat{P}_1$ that project into the singlet and triplet states respectively can 
be written as:
\begin{equation}
\hat{P}_0 = |\alpha\beta\rangle_{0}\langle\alpha\beta|_{0}
\label{eq:P0}
\end{equation}
\begin{equation}
\hat{P}_1 = |\beta\beta\rangle\langle\beta\beta| + |\alpha\beta\rangle_{1}\langle\alpha\beta|_{1} 
+ |\alpha\alpha\rangle\langle\alpha\alpha|
\label{eq:P1}
\end{equation}
where 
\begin{equation}
|\alpha\beta\rangle_{0} = \frac{1}{\sqrt{2}}\{\alpha(1)\beta(2) - \beta(1)\alpha(2)\},
\label{eq:singlet}
\end{equation}
\begin{equation}
|\alpha\beta\rangle_{1} = \frac{1}{\sqrt{2}}\{\alpha(1)\beta(2) + \beta(1)\alpha(2)\}.
\label{eq:triplet}
\end{equation}
we used the Slater geminal correlation factor\cite{TenNo:2004}, $f(r_{12}) = 
(1-\text{exp}(-\gamma r_{12})/\gamma)$, approximated with a fit to 6 Gaussian geminals\cite{Klopper:2005}. 
Optimized (empirical) values of $\gamma$ for different orbital basis sets are available in 
the literature\cite{Manby:2005,Klopper:2005,Peterson:2008}.\\\\
Finally, the second-order F12 correction energy is obtained from Eq.~(\ref{eq:Hylleraas})
by substituting the first-order explicitly correlated wave function,
\begin{equation}
E_{{\text{(2)}}_{\overline{\text{F12}}}} = \sum_{i<j} 2\tilde{V}^{ij}_{ij} + \tilde{B}^{ij}_{ij}.
\label{eq:F12 energy}
\end{equation}
The intermediates $\tilde{V}^{ij}_{ij}$ come from the $\langle\bar{0}|{\hat{H}}^{(1)}|1_{\text{F12}}\rangle$ and
$\langle1_{\text{F12}}|{\hat{H}}^{(1)}|\bar{0}\rangle$ terms of the Eq.~(\ref{eq:Hylleraas}) with $\hat{H}^{(1)} = \bar{H}$,
and thus depend on the $\hat{T}$ amplitudes,
\begin{equation}
\tilde{V}^{ij}_{ij} \equiv V^{ij}_{ij} + \frac{1}{2}(V^{ij}_{a_{ij}b_{ij}} + 
C^{ij}_{a_{ij}b_{ij}}) t^{a_{ij}b_{ij}}_{ij} + 
V^{ij}_{ia_j}t^{a_j}_{j} + V^{ij}_{a_ij}t^{a_i}_{i},
\label{eq:Vtilde}
\end{equation}
\begin{equation}
V^{ij}_{p_{ij}q_{ij}} \equiv \frac{1}{2} \bar{R}^{\thinspace ij}_{\alpha_{ij}\beta_{ij}} 
\bar{g}^{\thinspace \alpha_{ij}\beta_{ij}}_{p_{ij}q_{ij}},
\label{eq:V}
\end{equation}
\begin{equation}
C^{ij}_{a_{ij}b_{ij}} \equiv  F^{\alpha_{ij}}_{a_{ij}}\bar{R}^{\thinspace ij}_{\alpha_{ij}b_{ij}} 
+ F^{\alpha_{ij}}_{b_{ij}}\bar{R}^{\thinspace ij}_{a_{ij}\alpha_{ij}},
\label{eq:C}
\end{equation}
where $p_{ij},q_{ij}$ refer to the orbitals of the orbital basis set (OBS) (unoccupied + occupied) corresponding to pair $ij$.
The intermediate $\tilde{B}^{ij}_{ij}$ appears as a result of the resolution of
$\langle1_{\text{F12}}|{\hat{H}}^{(0)}|1_{\text{F12}}\rangle$ term with $\hat{H}^{(0)} = \hat{F}$,
\begin{equation}
\tilde{B}^{\thinspace ij}_{ij} \equiv B^{ij}_{ij} - X^{ik}_{ij}F^j_k - X^{kj}_{ij}F^i_k,
\label{eq:Btilde}
\end{equation}
\begin{equation}
B^{ij}_{ij} \equiv \bar{R}^{\thinspace ij}_{\alpha_{ij}\beta_{ij}} 
F^{\beta_{ij}}_{\gamma_{ij}}\bar{R}_{\thinspace ij}^{\alpha_{ij}\gamma_{ij}},
\label{eq:B}
\end{equation}
\begin{equation}
X^{ij}_{ij} \equiv \frac{1}{2} \bar{R}^{\thinspace ij}_{\alpha_{ij}\beta_{ij}} \bar{R}^{\thinspace \alpha_{ij}\beta_{ij}}_{ij}.
\label{eq:X}
\end{equation}
The excitations onto the full unoccupied space ($\alpha,\beta$) are taken into account through the addition 
of a new complementary auxiliary basis set (CABS)\cite{Valeev:2004ep}. From Eqs. (\ref{eq:Btilde}) and (\ref{eq:X}),
\begin{equation}
\tilde{B}^{\thinspace ij}_{ij} \leftarrow X^{ik}_{ij}F^j_k \equiv \frac{1}{2} \bar{R}^{\thinspace ik}_{\alpha_{ik}\beta_{ik}} \bar{R}^{\thinspace \alpha_{ij}\beta_{ij}}_{ij},
\label{eq:XB}
\end{equation}
construction of the $\tilde{B}^{ij}_{ij}$ intermediate involves coupling of the $ij$ pair with all other pairs $ik$
through the corresponding PNO overlap 
matrices. However, consistent with our earlier work on the LPNO-CCSD-F12 method\cite{Pavosevic:2014}, the following approximation is invoked
(to decouple the pairs),
\begin{equation}
X^{ik}_{ij}F^j_k \approx \frac{1}{2} \bar{R}^{\thinspace ik}_{\alpha_{ij}\beta_{ij}} \bar{R}^{\thinspace \alpha_{ij}\beta_{ij}}_{ij} F^j_k \equiv X^{ij_{F}}_{ij},
\label{eq:XAppr}
\end{equation}
where $j_{F}$ refers to the Fock-transformed occupied orbitals obtained through linear transformation of occupied orbitals with the 
occupied-occupied block of the Fock matrix. For the evaluation of the intermediate $B$, two 
approximations were employed: a) the D approximation which avoids the computation of the CABS-CABS block of 
the exchange matrix without the addition of any significant errors\cite{Pavosevic:2016bc}
b) the standard approximation (SA) which drops the CABS terms appearing in 
the contraction of the $V$ intermediate with $T$ amplitudes which only introduces errors
smaller than the residual basis set incompleteness error (BSIE)\cite{Valeev:2008,Torheyden:2008ev,Zhang:2012it}.\\\\
Finally, the spin-adaptation of Eq.~(\ref{eq:F12 energy}) yields\cite{Zhang:2012it}:
\begin{eqnarray*}
E_{\text{{(2)}}_{\overline{\text{F12}}}} &= &\sum_{I<J}\epsilon^{(2)}_{IJ} 
+\sum_{I,\bar{J}}\epsilon^{(2)}_{I\bar{J}} 
+ \sum_{\bar{I}<\bar{J}}\epsilon^{(2)}_{\bar{I}\bar{J}},\\
\label{eq:F12energy_sf}
\epsilon^{(2)}_{IJ} &= &\frac{1}{2}\tilde{V}^{IJ}_{IJ} +  \frac{1}{16}\tilde{B}^{IJ}_{IJ},\\
\label{eq:F12energy_sf_aa}
\epsilon^{(2)}_{I\bar{J}} &= &\frac{3}{4}\tilde{V}^{I\bar{J}}_{I\bar{J}} +  \frac{1}{4}\tilde{V}^{I\bar{J}}_{\bar{J}I}
+ \frac{9}{64}\tilde{B}^{I\bar{J}}_{I\bar{J}} +  \frac{3}{64}\tilde{B}^{I\bar{J}}_{\bar{J}I}
+ \frac{3}{64}\tilde{B}^{\bar{J}I}_{I\bar{J}} +  \frac{1}{64}\tilde{B}^{\bar{J}I}_{\bar{J}I},\\
\label{eq:F12energy_sf_ab}
\epsilon^{(2)}_{\bar{I}\bar{J}} &= &\frac{1}{2}\tilde{V}^{\bar{I}\bar{J}}_{\bar{I}\bar{J}} +  
\frac{1}{16}\tilde{B}^{\bar{I}\bar{J}}_{\bar{I}\bar{J}}.
\label{eq:F12energy_sf_bb}
\end{eqnarray*}
In the above equation, for the sake of generality, spin dependence of spatial orbitals 
is assumed: $I$ and $\bar{I}$ refer to the $\alpha$ and $\beta$ spatial orbitals of orbital $i$ 
respectively. The prefactors in these terms arise from the cusp coefficients ($\frac{1}{2}$ 
for singlet and $\frac{1}{4}$ for triplet). We have derived the pair-specific expressions of the F12 intermediates from the 
canonical equations  provided in the supplementary material of our earlier work\cite{Zhang:2012it}. Since this work is based on 
ROHF-like reference  determinants, spatial orbitals for both $\alpha$ and $\beta$ spins are identical. 
While DOMOs appear in both the spins as occupied orbitals, SOMOs appear as occupied orbitals for
the $\alpha$ spin, and as virtual orbitals for the $\beta$ spin. In this work, we have implemented a UHF (unrestricted HF) 
style F12 correction for all the pairs. Thus, F12 corrections for a given DOMO-DOMO pair involve 
calculations over three spin cases ($\alpha-\alpha$, $\alpha-\beta$, $\beta-\beta$) while SOMO-DOMO and SOMO-SOMO pairs 
required two ($\alpha-\alpha$, $\alpha-\beta$) and one ($\alpha-\alpha$) spin evaluations respectively.  
Efficient reduced-scaling evaluation of the F12 intermediates within the
\textit{SparseMaps} framework is described next\cite{Pinski:2015ii}. 

\subsection{The DLPNO Approximations}
\subsubsection{Orbital domains}
All occupied molecular orbitals must be localized to minimize their spatial extent.
For closed- and open-shell species the canonical Hartree-Fock and quasi restricted orbitals (QROs)\cite{Neese:2006} were used as the input orbitals for the Foster-Boys\cite{Foster:1960gk} localization algorithm.
It should be noted that DOMOs and SOMOs must be localized separately.

Projected atomic orbitals (PAOs) were used as an intermediate basis for the unoccupied orbital space.
The PAOs, \{$\tilde{\mu}$\},
are obtained by projecting out the contribution of occupied orbitals, \{$i$\}, from the atomic orbitals, $\{\mu\}$,
\begin{equation}
|\tilde{\mu}\rangle=\left(1-\sum_{i}|i\rangle\langle i|\right)|\mu\rangle.
\label{eq:PAOs}
\end{equation}
The PAOs are generally more localized than their canonical counterparts, and only a spatially-compact
subset of them is 
required to accurately describe the correlation amplitudes associated with a given (localized) occupied orbital
(or a set of such orbitals).
For a given occupied orbital $i$, only those PAOs are chosen are in its domain whose differential 
overlap integral (DOI)\cite{Pinski:2015ii} with $i$ is greater than the $T_{\text{CutDO}}$ threshold,
\begin{equation}
\mathrm{DOI}_{\thinspace i \tilde{\mu}} \equiv \sqrt{\int d \mathbf{x}\left|\phi_{i}
(\mathbf{x})\right|^{2}\left|\phi_{\tilde{\mu}}(\mathbf{x})\right|^{2}}.
\label{eq:DOI}
\end{equation}
The DOI is a measure of ``distance''
in Hilbert space between the orbital densities, $\left|\phi_{i}(\mathbf{x})\right|^{2}$ and 
$\left|\phi_{\tilde{\mu}}(\mathbf{x})\right|^{2}$; it can be efficiently evaluated in a linear scaling fashion 
using standard quadrature techniques\cite{Pinski:2015ii}.
The PAO domains are ``atom-complete'' in the sense that if
one or more of an atom's PAO is included in a given domain, all other PAOs associated with that atom are added
as well. Consequently, pair, triples, etc. domains are constructed from the union of the individual domains.

Both localized MOs or PAOs are sparsified further
to neglect contributions from AOs with coefficient less than
threshold $T_{\text{CutC}}$; this sparsification is also atom-complete, i.e. if an AO contributes
with coefficient greater than $T_{\text{CutC}}$ then contributions from all other AOs of that atom are included.
Of course, the AOs need to be normalized first in order to take advantage of this sparsity.

For the DLPNO-F12 procedure to be linear scaling the domains for occupied and
CABS AOs must also be introduced. These are defined as follows\cite{Pavosevic:2016bc}.
For a given orbital $i$ the domain of CABS AOs
includes all atoms which have one or more PAO with DOI to $i$ of not less than $T_{\text{CutDO}}/10$.
(note that we did not use PAO-like representation for the CABS basis). The CABS domain for pairs are obtained 
by taking a union of the individual CABS AO domains. The domain of occupied orbitals (including core) for a given 
{\em active} occupied orbital $i$ includes all orbitals with DOI to that orbital of not less than $T_{\text{CutDO}}/10$.
Thus, the domain sizes of occupied, PAO, and CABS AOs are controlled by a single parameter $T_{\text{CutDO}}$;
as it is lowered the domains grow roughly logarithmically.

The AO domains for local density fitting were defined as in previous work\cite{Pinski:2015ii}.

\subsubsection{Pair-natural orbitals}
PNOs are much more compact than PAOs and are hence employed as the 
final representation of the unoccupied space. 
In this work, PNOs are obtained by the diagonalization of the NEVPT2 pair-densities constructed in the PAO 
space and (as before) are truncated based on the parameter $T_{\text{CutPNO}}$.\cite{Saitow:2017bo} The choice of the NEVPT ansatz ensures that 
(1) a single set of PNOs (NEV-PNOs) is obtained which converges to the MP1-PNOs 
at the closed-shell limit since all the excitations involving the SOMOs disappear; (2) the wave function 
is intruder-state free as the zeroth-order wave function (Dyall's model Hamiltonian\cite{Dyall:1995}), unlike the 
Fock operator, includes the complete two-body interaction in the SOMO space; (3) the pair energies
and the pair wave function are invariant to the unitary transformations within the DOMO, SOMO and (pure) unoccupied spaces.

Thus, the use of NEV-PNOs offers significant advantages over the PNOs generated from UMP1 pair-densities\cite{Hansen:2011} 
which can suffer from the intruder-state problem. Also, in such an approach, three sets of PNOs are generated, resulting in 
(at least three times as many) integral transformations from the PAO to PNO space. Furthermore, it is quite difficult 
to match the closed-shell results with RHF based DLPNO implementations even after using the same $T_{\text{CutPNO}}$ 
values since the densities of the same-spin and opposite-spin pairs would have different truncations. This
often results in an unbalanced treatment of closed- and open-shell states, leading to errors in reaction energies, 
singlet-triplet gaps etc. Finally, even at zero PNO truncation, the
correlation energies obtained doesn't match with the canonical values: for diagonal pairs
of the same spin, PNOs cannot be generated (zero density), resulting in the removal of terms that depend on them
in the CCSD residual equations\cite{Saitow:2017bo}. 


\subsubsection{Pair-screenings}
The CCSD(T)$_{\overline{\text{F12}}}$ energy evaluation can be made linear-scaling 
only if the number of correlated electron clusters (pairs, triples, etc.) grows linearly with the system size.
In this regard, a hierarchy of approximations have been employed, beginning with the {\em pair-prescreening} step, where 
a pair $ij$ is neglected if DOI$_{\text{ij}} < T_{\text{CutDOij}}$ and dipole-approximated semi-local
NEVPT2 energy ($E_{\text{PreScr}}$) \cite{Guo:2016DLPNONEVPT2} is smaller than $T_{\text{CutPre}}$. $E_{\text{PreScr}}$ 
for that pair is added to the correlation energy. Consistent with the RHF implementation, a two-step guess procedure ({\em crude} and {\em fine}) is 
then followed. The domains in the {\em crude} guess step are chosen to be smaller than the ones used in the {\em fine} guess stage by 
using scaled versions of the {\em fine} guess truncation parameters, $T_{\text{CutDO\_Crude}} = 2 \times T_{\text{CutDO}}$, 
$T_{\text{CutMKN\_Crude}} = 10 \times T_{\text{CutMKN}}$. The pairs are classified on the basis of their correlation energies 
($\epsilon_p$) as {\em Crude\_CC} ($\epsilon_p > T_{\text{CutPairs}}$), {\em Crude\_PT2} 
($\epsilon_p > T_{\text{CutPairs\_MP2}}$) and {\em weak} pairs (remaining). Both {\em Crude\_CC} and {\em Crude\_PT2} pairs enter the {\em fine} guess
stage, where the above process is repeated (albeit with larger domains) and a CC pair list is obtained with the energies of {\em weak} 
and {\em PT2} pairs added to the total correlation energy. This two-step guess construction minimizes the computation
of three-index integrals while maintaining the accuracy of the correlation energies. It should also be noted
that pairs involving SOMOs do not take part in the {\em pre-screening} and {\em crude} guess steps due to large truncation errors\cite{Saitow:2017bo}.
Finally, PNOs are generated and then truncated for the CC pairs followed by the addition of a PNO truncation correction 
obtained from the difference between semi-local NEVPT2 correlation energies in the PAO and (truncated) NEV-PNO representations. 
Finally, all the integrals are transformed to the truncated PNO basis and CC residual equations are solved. 

Since the explicit correlation treatment is expected to be important for only spatially-close pairs, 
just like in the closed-shell formalism\cite{Pavosevic:2016bc} only a subset of the CC pairs ($ij$), defined by the criterion, DOI$_{ij} > 30 \times T_{\text{CutDOij}}$ are treated as the F12 pairs. 
Furthermore, {\em fine} guess maps were used for defining the domain sizes used in the F12 specific integral transformations.

\subsubsection{SOMO handling}
Since SOMOs appear as both occupied ($\alpha$ spin) and unoccupied orbitals ($\beta$ spin), they need to be included in 
the occupied, PAO, and PNO domains for the integral transformations. In this work, all $N_{\rm SOMO}$ SOMOs are included in every 
PAO domain of the occupied orbitals and the PAO-PNO transformation matrix for every pair is augmented by an $N_{\rm SOMO}$ by $N_{\rm SOMO}$ identity matrix. 
This avoids the {\em integral-direct} transformations of a number of integral classes which differ only by zero or more SOMOs: for example, 
the $(I\bar{A}_{IJ}|J\bar{B}_{IJ})$ integral batch, in which SOMOs appear in every orbitals space,
also contains other required 
integral batches, such as $(IA_{IJ}|JB_{IJ})$, $(\bar{I}\bar{A}_{\bar{I}\bar{J}}|\bar{J}\bar{B}_{\bar{I}\bar{J}})$, 
$(\bar{I}A_{\bar{I}\bar{J}}|\bar{J}B_{\bar{I}\bar{J}})$, etc. This strategy enables the re-use of 
optimized integral transformation routines of the closed-shell DLPNO infrastructure. Replicating SOMOs in different integral 
classes and PNOs of every pair does introduce some redundancy which can become a problem if the number of SOMOs is large.
Also, each SOMO shares the same orbital domain formed by taking an union of the domains of individual SOMOs
since the NEVPT2 energy expression of pairs (in the PAO basis) with SOMOs include terms with summations over all SOMO-SOMO pairs
\cite{Saitow:2017bo}.

As discussed before, the DLPNO-F12 procedure requires the construction of occupied-occupied domains. 
We make sure that all the SOMOs are included in each of these domains. This ensures that the
local OBS space for a pair i.e. local occupied orbitals + PNOs, become independent of the spin, making the local 
CABS space which is orthogonal to the local OBS space spin-independent as well.

\subsubsection{F12 integral transformations} 
Integral transformations are one of the most compute-intensive parts of a perturbative
F12 correction model like DLPNO-CCSD(T)$_{\overline{\text{F12}}}$.
Since our F12 calculations utilize two spaces (OBS, CABS) and 5 integral kernels: $\frac{1}{r_{12}}, f_{12}, f^2_{12}, 
\frac{f_{12}}{r_{12}}, [f_{12},[\hat{T},f_{12}]]$, {\em ten} sets of 3-index 2-electron integrals $(i p_{ij}|K|V_{ij})$ need
to be evaluated, with $K$ one of the 5 integral kernels and $p$ either the OBS or CABS space index. 
In addition, $(p_{ij}a_{ij}|\frac{1}{r_{12}}|V_{ij})$ integrals (where $a_{ij}$ is the PAO index), 
Fock-transformed (kernel: $f_{12}$; space: OBS, CABS; spin: $\alpha, \beta$) and $t_1$ transformed\cite{Pavosevic:2017kb} 
(kernel: $\frac{1}{r_{12}}, \frac{f_{12}}{r_{12}}$; space: OBS; spin: $\alpha, \beta$) integrals
are also required, resulting in a total of \textbf{19} 3-index integral transformations.

Evaluation of these 3-index
integrals is mostly identical to the closed-shell DLPNO F12 formalism\cite{Pavosevic:2016bc,Pavosevic:2017kb}
using the pair-extended SparseMaps.\cite{Pinski:2015ii}
Namely, the domain of an occupied orbital $i$ is augmented
with the domains of all those occupied orbitals $j$, if the pair $ij$ exists in the F12 pair list. Consequently,
for a given pair $ij$, integrals of $i$ and $j$ are sieved out from these pair-extended integrals and recombined through local density fitting,
followed by their transformation to the PNO basis to evaluate $ij$-specific intermediates. 

However, evaluation of the 
$(p_{ij}a_{ij}|\frac{1}{r_{12}}|V_{ij})$ integrals was significantly improved compared to the
closed-shell DLPNO-F12 formalism as follows.
As is the case for all 3-index integrals in the DLPNO framework,\cite{Pinski:2015ii} integral-direct transformation
is driven by the density-fitted atomic orbitals $X$:
for each batch of $X$ a set of integrals $(\mu_X\nu_X|X)$ is computed and then transformed to the PAO basis.
For the $(p_{ij}a_{ij}|\frac{1}{r_{12}}|V_{ij})$ integral evaluation in the closed-shell DLPNO CC-F12 formalism\cite{Pavosevic:2017kb}
the sets of AOs $\mu_X$ were calculated by the following composition of 
sparse maps:\cite{Pinski:2015ii} $\mathbf{L}(X\rightarrow p)\subset\mathbf{L}(p\rightarrow \mu)$ where $\mathbf{L}(X\rightarrow p) \equiv \mathbf{L}(X\rightarrow i)
\subset\mathbf{L}(i\rightarrow j)\subset\mathbf{L}(j\rightarrow p)$, 
whereas $\nu_X$ was constructed as: $\mathbf{L}(X\rightarrow p) \subset\mathbf{L}(p\rightarrow a) \subset\mathbf{L}(a\rightarrow \nu)$
where $\mathbf{L}(p\rightarrow a) \equiv \mathbf{L}(p\rightarrow i)\subset\mathbf{L}(i\rightarrow j)\subset\mathbf{L}(j\rightarrow a)$. 
Here, $\subset$ refers to the ``chaining'' operation, for example, chaining of two maps $\mathbf{L_1}(f\rightarrow g)$ and $\mathbf{L_2}(g\rightarrow h)$
produces another map $\mathbf{L_3}(f\rightarrow h)$, where a given $h$ is included in map $L_3$ only when there is a $g$ in map $L_1$ that is connected to 
$h$ in $L_2$ map. In the data structure sense, a sparse map $L$ can be thought of as a {\tt vector} of {\tt vector}s which allows the individual elements of 
this structure to have different sizes based on the ``connectedness'' of that given orbital which is of course determined by metrics like DOI.
Such definition of the $\nu_X$ domain including the ``chain'' of $\mathbf{L}(X\rightarrow p)$ and $\mathbf{L}(p\rightarrow a)$ (pair-extended) maps turns out to be too conservative, leading to excessive storage and operational costs.
This problem was avoided in this work by evaluating the 
$\mathbf{L}(X\rightarrow a)$ maps in the same fashion as the auxiliary to OBS maps: $\mathbf{L}(X\rightarrow a) \equiv 
\mathbf{L}(X\rightarrow i) \subset\mathbf{L}(i\rightarrow j)\subset\mathbf{L}(j\rightarrow a)$.
Even though both approaches have linear cost in system size, the new algorithm has a significantly reduced prefactor without sacrificing the accuracy.

\section{Computational Details}\label{sec:CompDetails}
All numerical experiments were performed using a developmental version of ORCA 4.2. ROHF and quasi 
restricted orbitals (QROs) were used as the reference orbitals in UHF based computations.
Core orbitals were kept doubly occupied in all correlated computations.
Evaluation of the $B$ F12 intermediate followed the F12/D approximation.\cite{Pavosevic:2016bc}
A number of orbital basis sets (OBS) were utilized in this work: aug-cc-pV[X]Z and aug-cc-pV[X](+d)Z basis sets of Dunning et al.\cite{Woon93}, 
F12-optimized basis sets of Peterson et al.\cite{Peterson:2008}, cc-pV[X]Z-F12, with X = D,T, where X is the cardinal number of the basis set and 
the def2-family of basis sets developed by Weigend and Ahlrichs\cite{Weigend2005}, def2-TZVP, def2-TZVPP and def2-SVP. 
For a given OBS, the corresponding density-fitted basis set (DFBS) was used unless stated otherwise\cite{Weigend:2002,Weigend2006,Hattig:2005}, 
ex. for aug-cc-pVDZ basis set, aug-cc-pVDZ/RI was used as the auxiliary basis. For the cc-pV[X]Z-F12 basis sets, aug-cc-pV[X]Z/RI was used as the DFBS. 
In the construction of the CABS space, {\em uncontracted} def2-TZVPP and def2-QZVPP basis sets were used when 
the standard optimized cc-pV[X]Z-F12-OptRI\cite{Yosaf2008} basis set (X=D,T) family was not available for the given molecular species. 

The convergence of the correlation energy with respect to the DLPNO truncation parameters
($T_{\text{CutPNO}}$, $T_{\text{CutDO}}$, $T_{\text{CutPairs}}$, $T_{\text{CutMKN}}$) 
was assessed for the doublet trityl radical at the B3LYP/def2-TZVPP equilibrium geometry;
these computations utilized the \{aug-cc-pVDZ, cc-pVDZ-F12-OptRI, aug-cc-pV5Z-RI\}
\{OBS, CABS, DFBS\} triplet.
The Hartree-Fock orbitals were computed using the {\tt VeryTightSCF} SCF convergence settings of ORCA.
The computational scaling with size was assessed for
n-alkanes ranging from \ce{C20H42} to \ce{C160H322}
at idealized quasilinear geometry using \{def2-TZVP, cc-pVDZ-F12-OptRI, def2-TZVP/C\} basis set
triplet. In these studies the Coulomb operator was
approximated by the standard $\mathcal{O}(N^3)$
density-fitting  procedure using the def2/J basis set\cite{Weigend2006}, which we will refer to as 
the coulomb density-fitted basis (CDFBS), while the exchange operator was described by the 
linear scaling chain of spheres approximation (COSX)\cite{Izsak:2011} used with default grid parameters.

The accuracy of the present method was analyzed and compared with extrapolation based techniques 
in heats of formation calculations on a representative set of molecules out of the ANL database of 348 core
 combustion species\cite{Klippenstein:2017} and reaction energies of 50 open-shell reactions\cite{Knizia:2009}.
Finally, computational timings were reported for some chemically relevant 
medium and large sized molecules.

All the calculations reported in this work were carried out on multicore nodes (although we utilized only 4 cores)
with dual E5-2683v4 2.1GHz (Broadwell) and dual E5-2680v3 2.5GHz (Haswell) processors, each node with a total 
memory of 512 GB.

\section{Results}\label{sec:Results}
\subsection{Convergence of the CCSD$_{\overline{\text{F12}}}$ correlation energy with respect to the DLPNO truncation parameters}
To be usable in chemical applications reduced-scaling methods must allow robust control of precision, hence
the errors due to the sparsifying approximations should decay monotonically and rapidly as the
truncation parameters approach zero; when all truncation parameters vanish the reduced-scaling method
should become equivalent to the canonical counterpart.
To this end we performed a systematic analysis of the convergence 
behavior of the UHF-based DLPNO CCSD$_{\overline{\text{F12}}}$ 
correlation energies for the trityl radical in the doublet state, a medium-sized system used previously by some of us
to assess the open-shell DLPNO-CCSD method.\cite{Saitow:2017bo}
(the lack of support for density fitting in the open-shell {\em canonical}
CCSD code in ORCA prevented us from using a larger molecule).
To be able to compare the canonical and DLPNO results
we used the aug-cc-pV5Z-RI as DFBS to minimize the errors introduced due to the density-fitting 
approximation used in the DLPNO procedure.

\begin{figure}[ht] 
  \begin{subfigure}[b]{0.5\linewidth}
    \centering
    \includegraphics[width=0.85\linewidth]{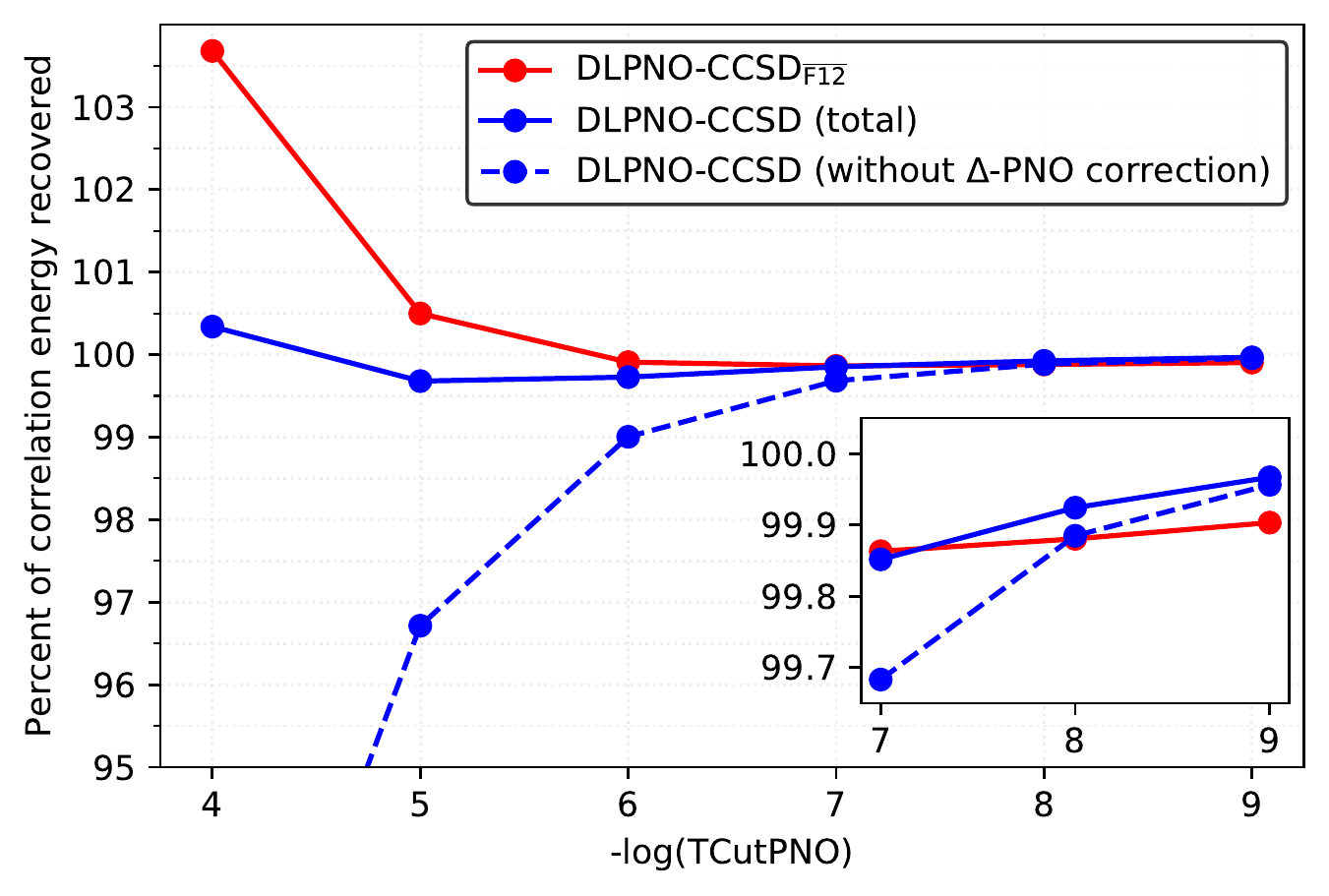} 
    \caption{$T_{\text{CutPNO}}$} 
    \label{fig1:a} 
  \end{subfigure}
  \begin{subfigure}[b]{0.5\linewidth}
    \centering
    \includegraphics[width=0.85\linewidth]{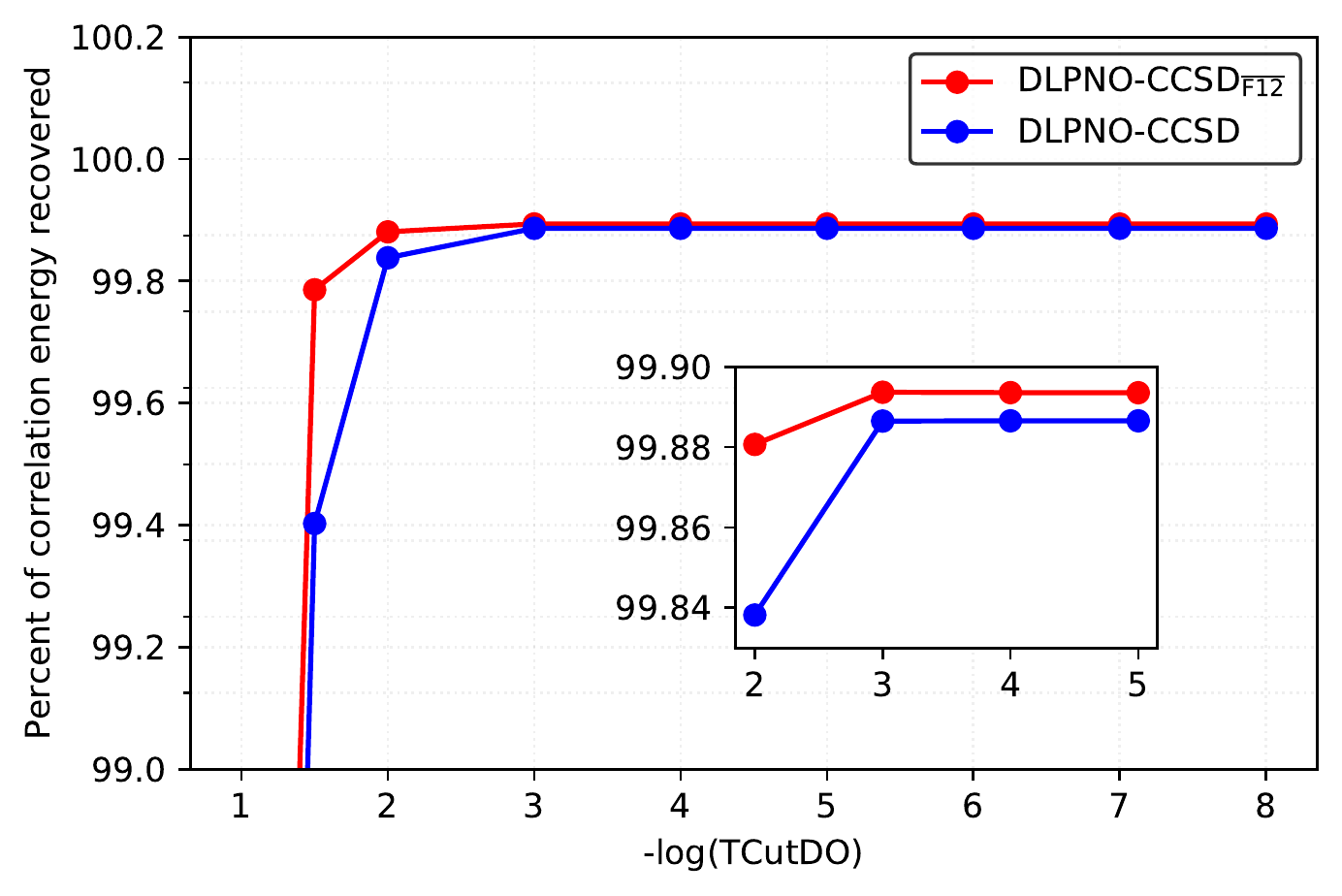} 
    \caption{$T_{\text{CutDO}}$} 
    \label{fig1:b} 
  \end{subfigure} 
  \begin{subfigure}[b]{0.5\linewidth}
    \centering
    \includegraphics[width=0.85\linewidth]{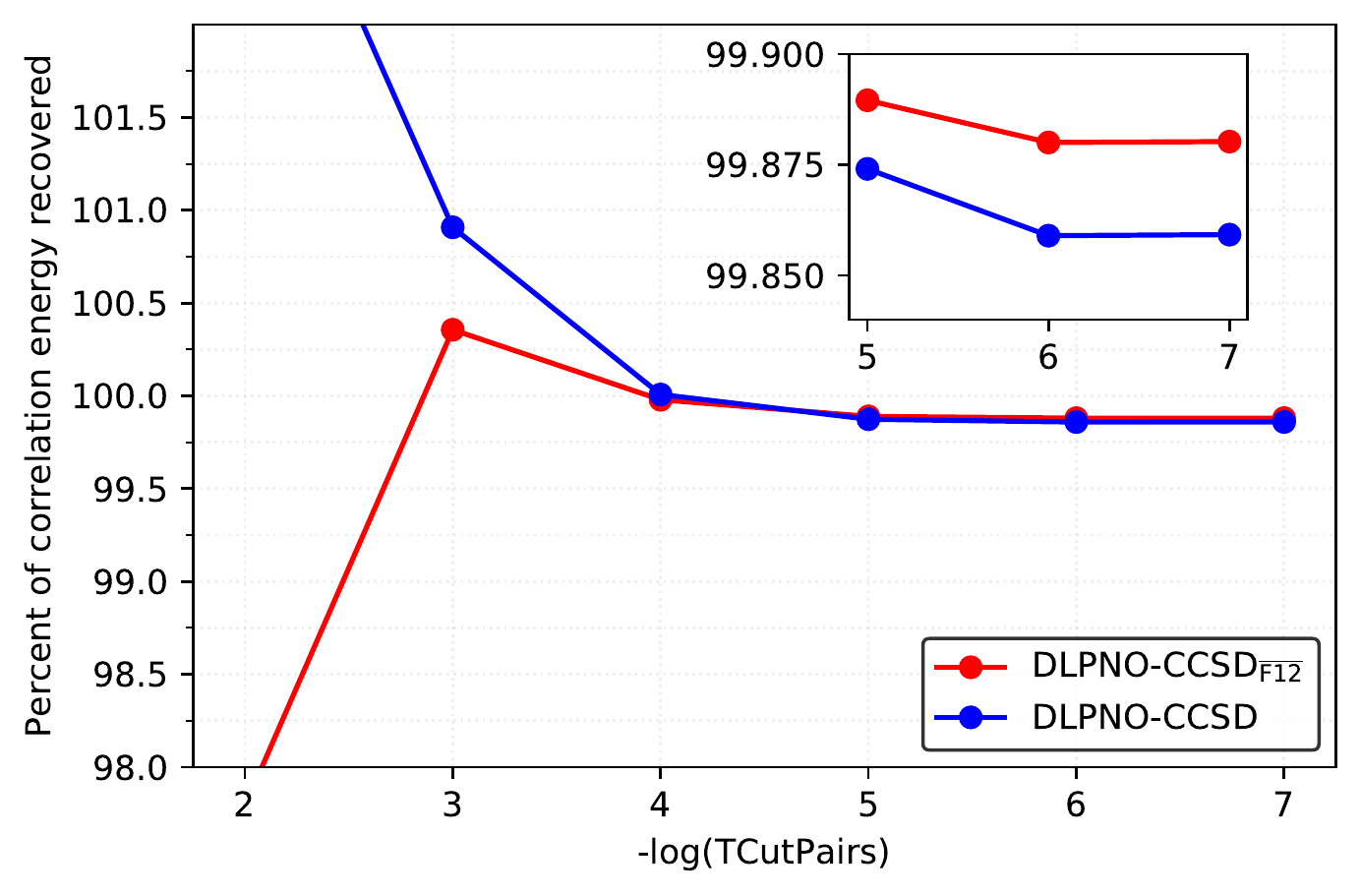} 
    \caption{$T_{\text{CutPairs}}$} 
    \label{fig1:c} 
  \end{subfigure}
  \begin{subfigure}[b]{0.5\linewidth}
    \centering
    \includegraphics[width=0.85\linewidth]{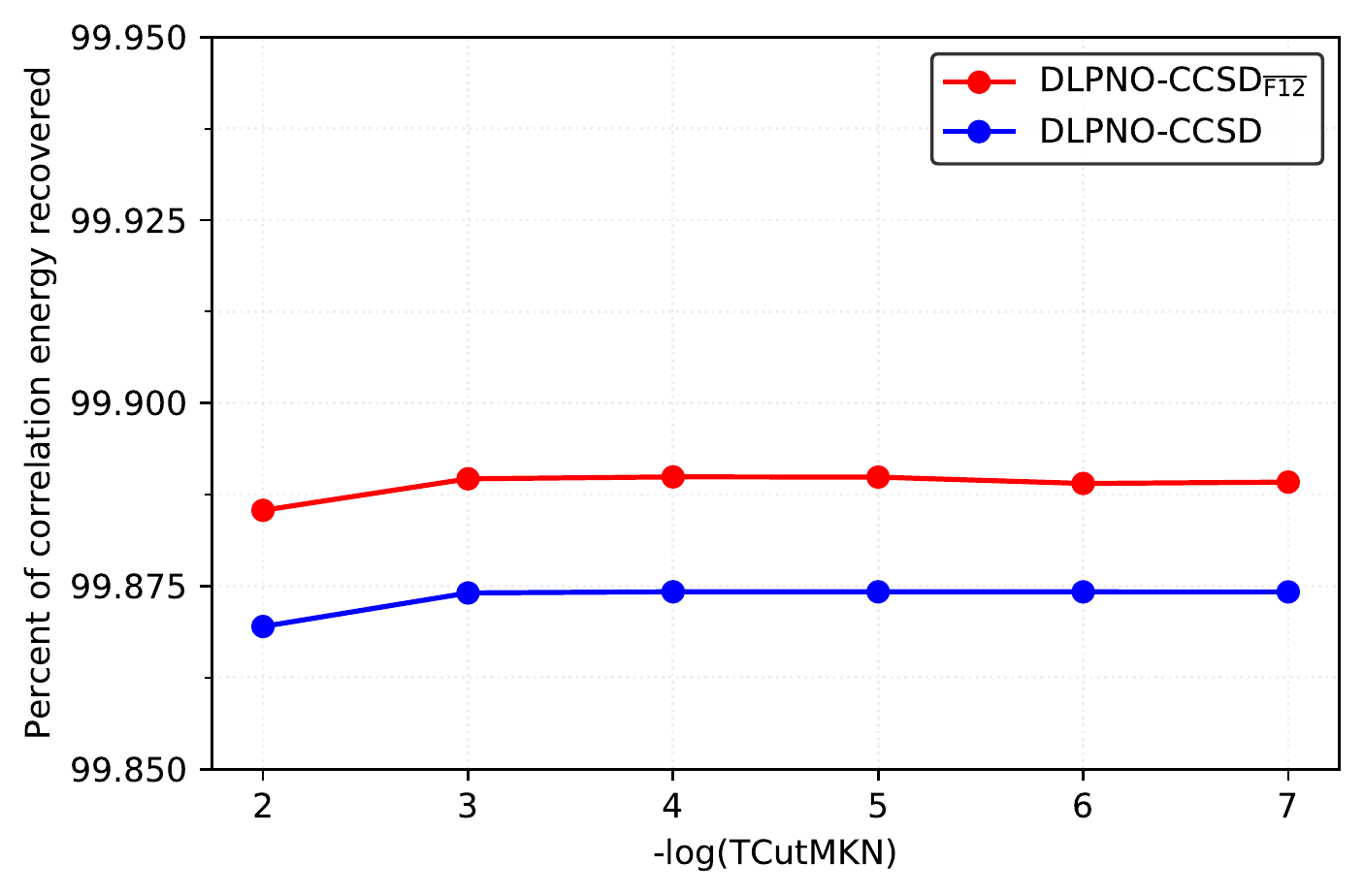} 
    \caption{$T_{\text{CutMKN}}$} 
    \label{fig1:d} 
  \end{subfigure} 
  \captionsetup{justification=raggedright,singlelinecheck=false}
  \caption{{\footnotesize DLPNO-CCSD and DLPNO-CCSD$_{\overline{\text{F12}}}$ valence correlation energies of doublet
trityl radical shown in terms of the percentage of the respective canonical energies, as
a function of the truncation parameters. All other truncation parameters are set to zero for (a) and
to {\tt TightPNO} settings: \{$T_{\text{CutPNO}}=10^{-7}$, $T_{\text{CutDO}}=5*10^{-3}$, $T_{\text{CutMKN}}=10^{-4}$, $T_{\text{CutPairs}}=10^{-5}$\} for (b), (c) and (d).}}
  \label{fig1} 
\end{figure}

Fig.~\ref{fig1:a} illustrates the convergence behavior of the DLPNO-CCSD (blue) and DLPNO-CCSD$_{\overline{\text{F12}}}$ 
(red) valence correlation energies of doublet trityl radical as a function of the PNO truncation 
parameter, $T_{\text{CutPNO}}$ when all other truncation parameters are set to 0. It can be seen that both 
curves converge monotonically towards their respective canonical correlation energies and at the PNO truncation 
of $10^{-7}$, around 99.86\% of the canonical energy is recovered for both the methods. However, the F12 energies show 
a slower convergence than its conventional counterpart: at a truncation of $10^{-9}$, DLPNO-CCSD recovers around 99.97\% 
while DLPNO-CCSD$_{\overline{\text{F12}}}$ yields a little more than 99.9\% of their canonical correlation energies. Also, at looser thresholds, 
the errors in the F12 energies are higher, which is consistent with the behavior that we observed in the 
LPNO-CCSD(2)$_{\overline{\rm F12}}$ method\cite{Pavosevic:2014}.
Note that the errors would be significantly larger in both conventional and F12 CCSD energies
if the PNO incompleteness correction ($\Delta-\text{PNO}$)\cite{Neese:2009db} were not included, as illustrated for
the CCSD case with the dashed line. Further reduction of the truncation error when $T_{\text{CutPNO}}\leq 10^{-8}$ is possible
by reoptimizing the PNOs for the case of the coupled-cluster wave function.\cite{Clement:2018dq}

Figure ~\ref{fig1:b} demonstrates the 
dependence of the DLPNO-CCSD and DLPNO-CCSD$_{\overline{\text{F12}}}$ energies on $T_{\text{CutDO}}$, 
the truncation parameter that defines the size of PAO domains. All other truncation parameters are set to {\tt TightPNO} settings: 
\{$T_{\text{CutPNO}}=10^{-7}$, $T_{\text{CutDO}}=5 \times 10^{-3}$, $T_{\text{CutMKN}}=10^{-4}$, $T_{\text{CutPairs}}=10^{-5}$\}. 
Unlike the PNO graph above, both the methods exhibit similar behavior and converge towards 
the canonical energy from below, with slightly lower errors in the F12 energies, in agreement with the findings of Werner\cite{Werner:2008} and our observations in 
the DLPNO-CCSD$_{\overline{\text{F12}}}$ method\cite{Pavosevic:2017kb} for closed-shell systems. At $T_{\text{CutDO}}=10^{-2}$, the \% recovery for DLPNO-CCSD and DLPNO-CCSD$_{\overline{\text{F12}}}$ energies 
are 99.84 and 99.88 respectively while at at $T_{\text{CutDO}}=10^{-3}$, both curves have essentially converged to $\approx99.89\%$. 

Similar convergence behavior is observed for $T_{\text{CutPairs}}$ (Fig.\ref{fig1:c}) and $T_{\text{CutMKN}}$ (Fig.\ref{fig1:d}) truncation parameters 
with F12 energies recovering a slightly higher percentage of correlation energy near convergence.
In Fig.\ref{fig1:d}, the errors for both the methods are quite small even at $T_{\text{CutMKN}}=10^{-2}$, which is reflective of the 
large density-fitting basis set (aug-cc-pV5Z/RI) used in these calculations. On average $\approx 99.86\%$ of the canonical 
correlation energy is recovered at the {\tt TightPNO} setting.

\subsection{Accuracy studies}
\subsubsection{Heats of formation}
The total correlation energy of the doublet triphenyl radical is around 3 Hartrees and even with an
overall accuracy of 99.86\%, the total error comes to be around 2.7 kcal/mol, much higher than the desired 
chemical accuracy of 1 kcal/mol. However, most of the chemical phenomena actually depend on relative energies 
of some kind and not absolute energies. Thus, its important to study the effect of all the DLPNO truncations 
on the accuracy of relative energy calculations.
To this end, we assessed the accuracy of the DLPNO-CCSD(T)$_{\overline{\text{F12}}}$ method for calculating heats of formation 
of 50 largest compounds out of the ANL database of 344 core combustion species\cite{Klippenstein:2017}.
Following the works of Klippenstein and co-workers\cite{Klippenstein:2017}, we use \ce{H2}, \ce{CH4}, \ce{H2O} and \ce{NH3}
as the reference species for the \ce{H}, \ce{C}, \ce{O}, and \ce{N} elements respectively. Thus, for the formation of $\text{C}_a\text{O}_b\text{N}_c\text{H}_d$ species, the 
following ``working reaction'' is considered,
\begin{equation}
 a \mathrm{CH}_{4}+b \mathrm{H}_{2} \mathrm{O}+c \mathrm{NH}_{3}+d / 2 \mathrm{H}_{2} \rightarrow 
\mathrm{C}_{a} \mathrm{O}_{b} \mathrm{N}_{\mathrm{c}} \mathrm{H}_{d}+(2 a+b+3 c / 2) \mathrm{H}_{2},
\end{equation}
and the enthalpy of formation of $\text{C}_a\text{O}_b\text{N}_c\text{H}_d$ can be calculated as\footnote{The equation displayed in Ref\cite{Klippenstein:2017}has a small sign-related typo: The sign before the group with $E\left(\mathrm{C}_{a} \mathrm{O}_{b} \mathrm{N}_{c} \mathrm{H}_{d}\right)$ should be positive instead of negative.}
\begin{equation}
\begin{array}{l}{\Delta_{\mathrm{f}} H_{0}^{\circ}\left(\mathrm{C}_{a} \mathrm{O}_{b} \mathrm{N}_{\mathrm{c}} \mathrm{H}_{d}\right)} \\\\
{=\left[a \Delta_{\mathrm{f}} H_{0}^{\circ}\left(\mathrm{CH}_{4}\right)+b \Delta_{\mathrm{f}} H_{0}^{\circ}\left(\mathrm{H}_{2} \mathrm{O}\right)+c \Delta_{\mathrm{f}} H_{0}^{\circ}\left(\mathrm{NH}_{3}\right)\right.} \\\\
{\left.\quad+(d /2 -2a-b-3 c / 2) \Delta_{\mathrm{f}} H_{0}^{\circ}\left(\mathrm{H}_{2}\right)\right]} \\\\
{\quad+\left[E\left(\mathrm{C}_{a} \mathrm{O}_{b} \mathrm{N}_{c} \mathrm{H}_{d}\right)-a E\left(\mathrm{CH}_{4}\right)-b E\left(\mathrm{H}_{2} \mathrm{O}\right)\right.} \\\\ 
{\left.\quad-c \thinspace E\left(\mathrm{NH}_{3}\right)+(2 a+b+3 c / 2-d / 2) E\left(\mathrm{H}_{2}\right)\right]}.
\end{array}
\end{equation}
In this work, we are only interested in the contribution of the non-relativistic electronic energy to 
the heats of formation and $E$ in the above equation will refer to the same. We compare the heats of formation obtained using DLPNO-CCSD(T)$_{\overline{\text{F12}}}$ and 
CBS extrapolation of canonical CCSD(T) energies within the ANL0 scheme. The extrapolations employed aug$'$-cc-pVQZ (a$'$QZ) and aug$'$-cc-pV5Z (a$'$5Z) 
basis sets while using the following formula: $E_{CBS} = E_{a'nZ} + \alpha (E_{a'nZ} - E_{a'(n-1)Z})$.
In the primed basis sets (a$'$QZ), diffuse functions are omitted except the s functions on the H atoms
and the s and p functions on the C, N and O atoms. 
$\alpha=0.75$ was chosen such that the RMSD between (experimental) ATcT\cite{Branko:2004} values and the ANL0 predictions were minimized\cite{Klippenstein:2017}.
Also, the extrapolated results using optimized values of $\alpha$ 
were shown to be in close agreement with the 1/$l_{\text{max}}^{3.7}$ extrapolation scheme. 
The geometries of these 50 molecules were optimized with canonical CCSD(T)
using the cc-pVTZ basis set. The energies of 21 closed-shell species (out of 50) were
calculated by the RHF-DLPNO-CCSD(T) method. 
We have used four sets of basis set triplets, \{OBS, CABS, DFBS\}: 
\{cc-pVDZ-F12, cc-pVDZ-F12-OptRI, aug-cc-pVDZ/RI\}, \{cc-pVTZ-F12, cc-pVTZ-F12-OptRI, aug-cc-pVTZ/RI\}, \{aug-cc-pVDZ, cc-pVDZ-F12-OptRI, aug-cc-pVDZ/RI\} 
and \{aug-cc-pVTZ, cc-pVDZ-F12-OptRI, aug-cc-pVTZ/RI\} with the DLPNO-CCSD(T)$_{\overline{\text{F12}}}$ method
using both iterative, ({\tt T1}),\cite{Yang:2018,Yang:2020} and semi-canonical non-iterative, ({\tt T0}),\cite{Riplinger:2016} variants of the triples correction. 
From now on, we will refer to the cc-pV[X]Z-F12 and aug-cc-pV[X]Z basis sets as [X]Z-F12 and a[X]Z basis sets, respectively.

\begin{figure}[ht]
  \begin{subfigure}[b]{0.5\linewidth}
    \centering
    \includegraphics[width=0.85\linewidth]{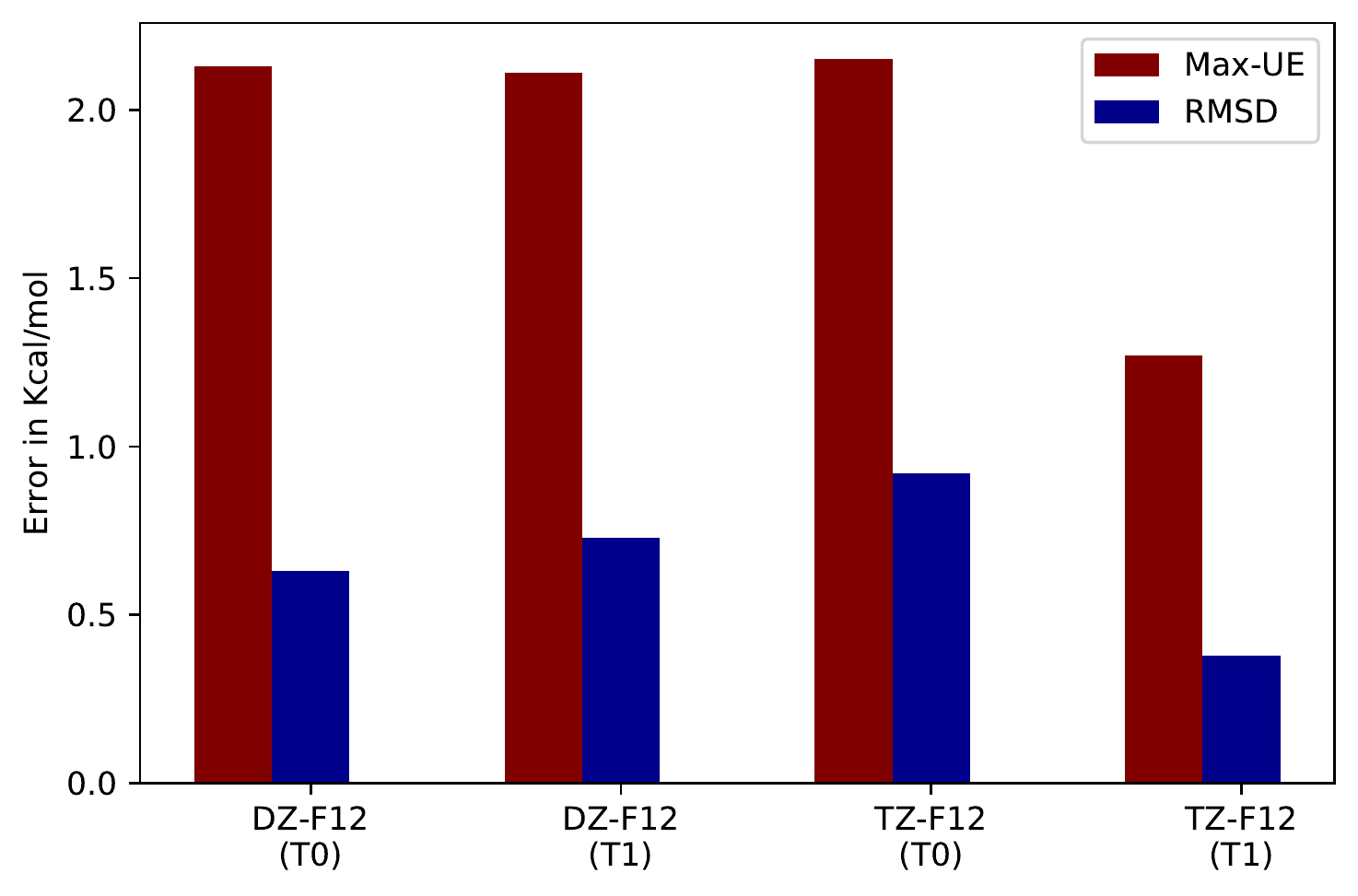}
    \caption{XZ-F12 basis}
    \label{fig3:a}
  \end{subfigure}
  \begin{subfigure}[b]{0.5\linewidth}
    \centering
    \includegraphics[width=0.85\linewidth]{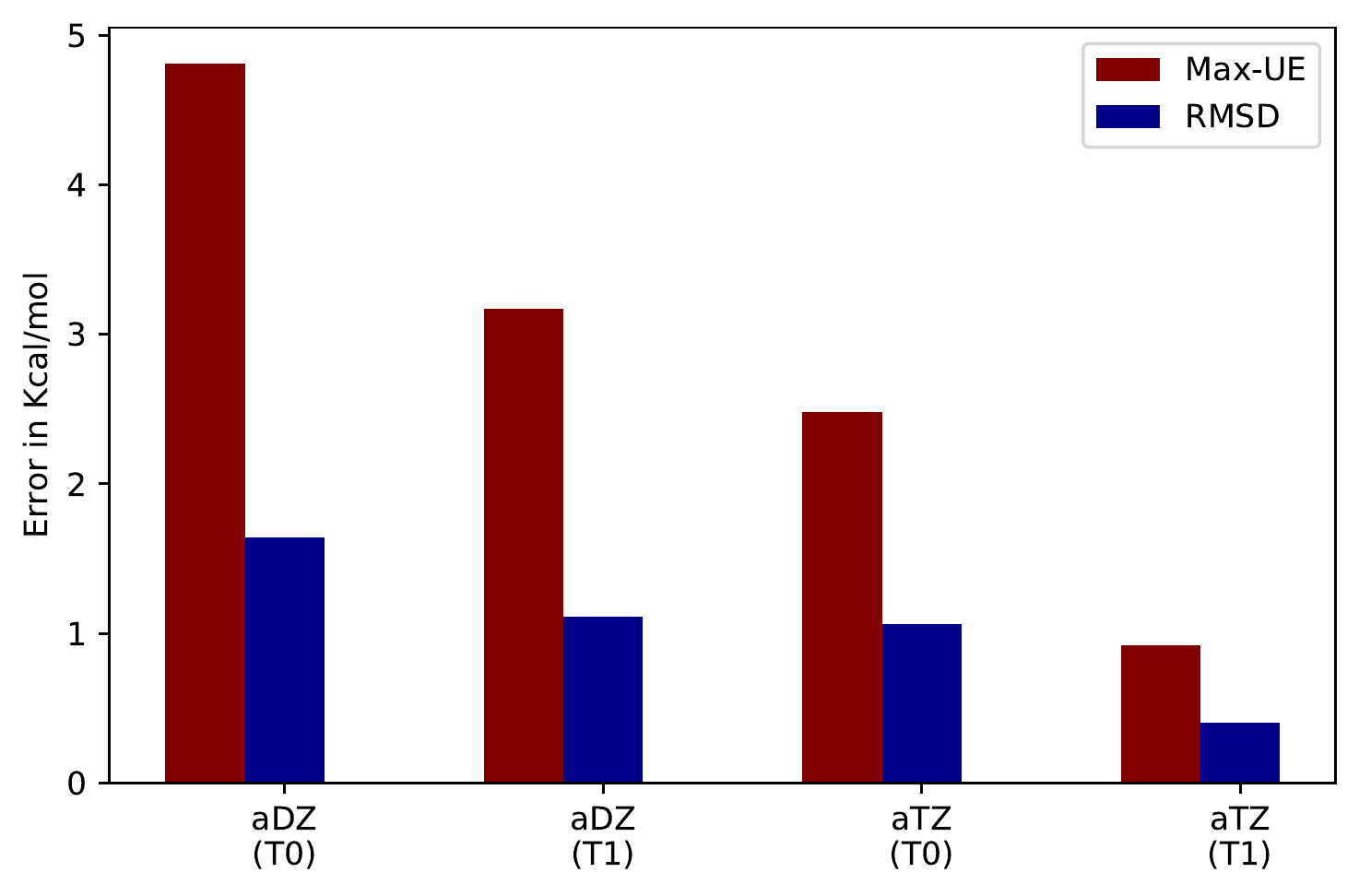}
    \caption{aXZ basis}
    \label{fig3:b}
  \end{subfigure}
  \captionsetup{justification=raggedright,singlelinecheck=false}
\caption{{\footnotesize Max unsigned error and RMSD of the UHF-DLPNO-CCSD(T)$_{\overline{\text{F12}}}$ method with 
          a) cc-pV[D,T]Z-F12 and b) aug-cc-pV[D,T] basis sets. Errors are calculated with respect to the 
          canonical CBS values using the a$'$5Z/a$'$QZ extrapolation technique of Ref.\cite{Klippenstein:2017}}.}
\label{fig3}
\end{figure}
The computed DLPNO-CCSD(T)$_{\overline{\text{F12}}}$ and the extrapolated CBS CCSD(T) heats of formation are tabulated in the supplementary information.
Consider first the DZ-F12 results (Fig.~\ref{fig3:a}) with the {\tt T0} approximation: the max and RMSD 
are -2.13 and 0.63 kcal/mol, respectively. Quite surprisingly, the 
more accurate T1 approximation gives higher RMSD for the same basis set: max and RMSD are 2.11 and 0.73 kcal/mol, respectively. With the larger TZ-F12 basis set 
the corresponding errors (max error, RMSD) for the {\tt T0} and {\tt T1} approximations are (-2.15, 0.92) and (-1.27, 0.38) kcal/mol, respectively. The errors in the 
smallest basis set used, aDZ (Fig.~\ref{fig3:b}) are the largest: (-4.81, 1.64) and (-3.17, 1.11) kcal/mol. On the other hand, the aTZ basis set yields errors of (-2.48, 1.06) 
and (-0.92, 0.40) kcal/mol for the {\tt T0} and {\tt T1} approximations, respectively. These results clearly indicate that the less expensive {\tt T0} approximation is unable to provide errors 
within the chemical accuracy and the more accurate {\tt T1} approximation should be preferred for these calculations which is consistent with the findings of Liakos et al.\cite{G.Liakos2019} 
Furthermore, the DZ-F12 basis set results with the {\tt T0} approximation is at best fortuitous since they 
are lower than the corresponding TZ-F12 and aTZ errors: (1.73, 0.45) vs (1.77, 0.69) and (2.11, 0.86) kcal/mol, respectively. The DLPNO-CCSD(T1)-F12 method 
with both TZ-F12 and aTZ basis sets yield RMSD of less than 0.40 kcal/mol at a fraction of the cost of the extrapolated canonical CCSD(T) procedure.
However, given the smaller size of the aTZ basis compared to that of TZ-F12 the use of the former
seems to be more economical in this case and is cautiously recommended for future applications; a more extensive benchmarking is however warranted.

\begin{figure}
\includegraphics[width=8cm]{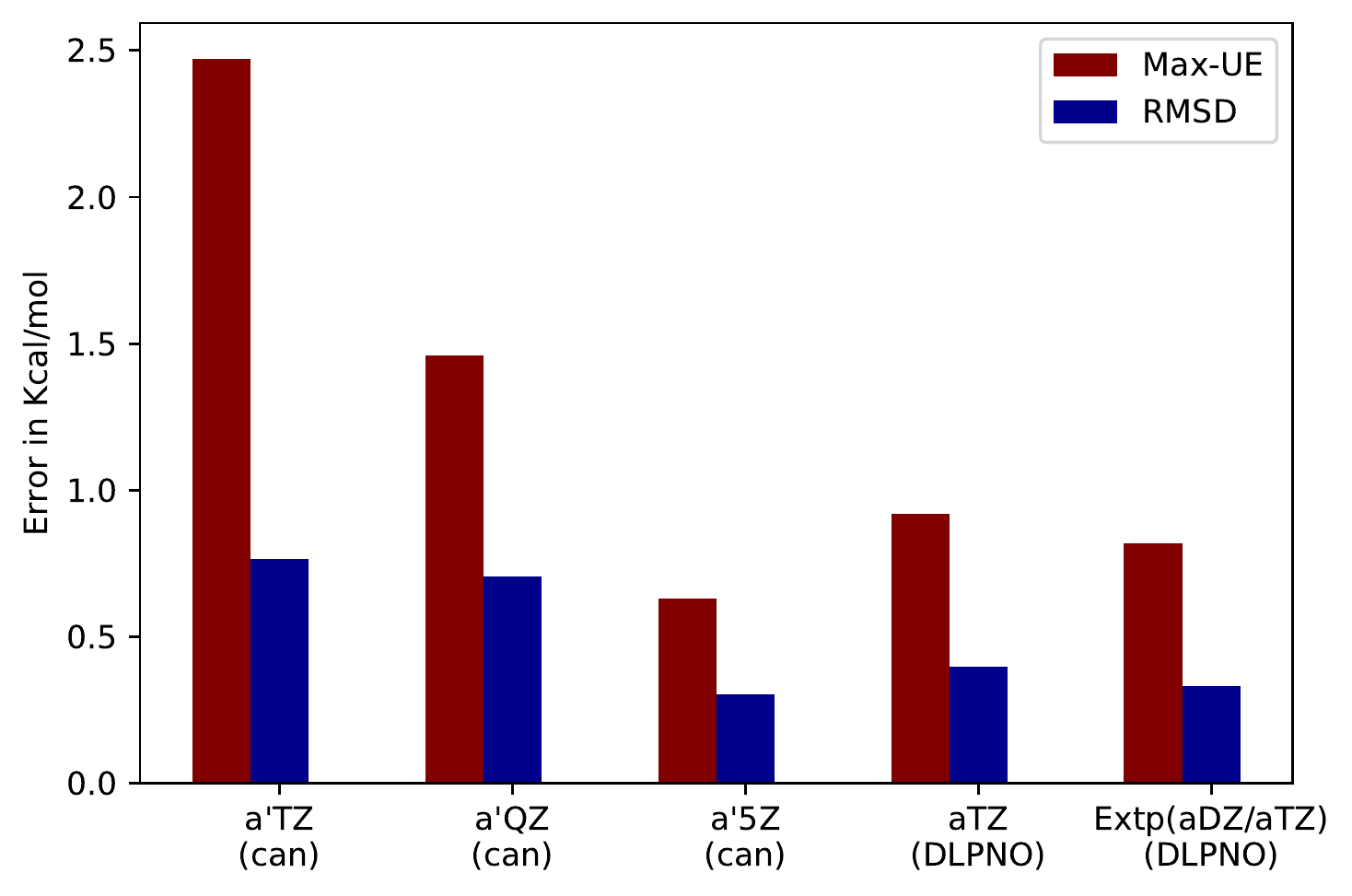}
\centering
\captionsetup{justification=raggedright,singlelinecheck=false}
\caption{{\footnotesize Max unsigned error and RMSD of canonical CCSD(T) method with a$'$TZ, a$'$QZ and a$'$5Z basis sets
and UHF-DLPNO-CCSD(T)$_{\overline{\text{F12}}}$ method with aTZ basis. Extrap refers to the extrapolated results with 1/$l_{\text{max}}^3$ scheme. Errors are calculated with 
respect to the canonical CBS values using the a$'$5Z/a$'$QZ extrapolation technique of Ref.\cite{Klippenstein:2017}}.}
\label{fig:extrapolate}
\end{figure}

Fig.~\ref{fig:extrapolate} compares the errors obtained from canonical CCSD(T) method using a$'$TZ, a$'$QZ and a$'$5Z basis sets 
with the DLPNO-CCSD(T)$_{\overline{\text{F12}}}$ method using the aTZ basis set. It can be seen that the DLPNO errors are only slightly worse than
the canonical a$'$5Z basis errors: (0.63, 0.30) vs (0.92, 0.40) kcal/mol. Furthermore, we also carried out the simple $X^{-3}$ based extrapolation 
scheme with the DLPNO-aDZ and DLPNO-aTZ F12 energies and the errors are further reduced to (0.82, 0.33) kcal/mol. 
\subsubsection{Reaction energies}
We chose the reaction energies (REs) of 50 ``open-shell'' reactions
originally devised by Knizia et al.\cite{Knizia:2009} as a more challenging benchmark for the DLPNO-CCSD(T)$_{\overline{\text{F12}}}$ method.
Out of the 47 molecules involved in these reactions, 17 are open-shell species, which include atoms, 2nd and 3rd period
elements in doublet and triplet spin states. Large reaction energies coupled with a strong basis set dependence  
makes this test set quite challenging for the F12 methods. The aug-cc-pVT(+d)Z (denoted for
simplicity as aVTZ) and aug-cc-pVQ(+d)Z (denoted as aVQZ) basis sets were employed 
as the OBS with cc-pVQZ-F12-OptRI and aug-cc-pVQZ/RI as the CABS and DFBS respectively. Furthermore, due to the higher accuracy requirements
and relatively small system size, the more accurate {\tt T1} variant of (T)
was used with {\tt TightPNO} and {\tt VeryTightPNO}
(\{$T_{\text{CutPNO}}=10^{-8}$, $T_{\text{CutDO}}=5*10^{-3}$, $T_{\text{CutMKN}}=10^{-4}$, $T_{\text{CutPairs}}=10^{-5}$\}) truncation settings.
ROHF orbitals were used in all of these calculations.

\begin{table} 
\tiny
\captionsetup{font=scriptsize,justification=raggedright,singlelinecheck=false}
\caption{Total DLPNO-CCSD(T)$_{\overline{\text{F12}}}$ correlation energies recovered (\%) with respect to the canonical values with aVTZ and 
         aVQZ basis sets at {\tt TightPNO} and {\tt VeryTightPNO} truncation settings. The numbers in the paranthesis 
         denote the errors in kcal/mol.}
\begin{ruledtabular}
\begin{tabular}{lcccccccc}
 {} &  \multicolumn{4}{c}{aVTZ} & \multicolumn{4}{c}{aVQZ}\\
\cmidrule(r){2-5}\cmidrule(l){6-9} 
 {} &  \multicolumn{2}{c}{\tt TightPNO}  & \multicolumn{2}{c}{\tt VeryTightPNO}  & \multicolumn{2}{c}{\tt TightPNO} & \multicolumn{2}{c}{\tt VeryTightPNO}\\
\bottomrule
{} & {} & {} & {} & {} & {} & {} & {} & {} \\

HCl    &   99.86   & (-0.23)&    99.97    &(-0.04)&    99.77    &(-0.37)&    99.93    &(-0.12)\\ 
Cl     &  100.11   & ( 0.15)&   100.24    &( 0.33)&    99.85    &(-0.20)&    99.99    &(-0.01)\\
H2     &   99.98   & (-0.01)&    99.98    &(-0.01)&    99.90    &(-0.03)&    99.90    &(-0.03)\\
F2     &   99.87   & (-0.49)&    99.97    &(-0.13)&    99.82    &(-0.70)&    99.91    &(-0.36)\\
H2O    &   99.96   & (-0.08)&    99.97    &(-0.05)&    99.90    &(-0.20)&    99.92    &(-0.15)\\
HF     &   99.96   & (-0.07)&    99.98    &(-0.03)&    99.88    &(-0.23)&    99.93    &(-0.13)\\
O      &  100.39   & ( 0.46)&   100.40    &( 0.47)&   100.05    &( 0.06)&   100.05    &( 0.06)\\
CH4    &   99.97   & (-0.04)&    99.98    &(-0.03)&    99.91    &(-0.14)&    99.91    &(-0.13)\\
OH     &  100.18   & ( 0.27)&   100.16    &( 0.24)&   100.00    &( 0.01)&   100.01    &( 0.01)\\
CH3    &   99.98   & (-0.02)&    99.99    &(-0.02)&    99.95    &(-0.06)&    99.96    &(-0.06)\\
CO     &   99.93   & (-0.17)&    99.99    &(-0.03)&    99.86    &(-0.35)&    99.94    &(-0.16)\\
CO2    &   99.89   & (-0.46)&    99.97    &(-0.12)&    99.83    &(-0.73)&    99.92    &(-0.36)\\
Cl2    &   99.76   & (-0.71)&    99.95    &(-0.15)&    99.73    &(-0.84)&    99.91    &(-0.29)\\
CH3Cl  &   99.87   & (-0.38)&    99.96    &(-0.11)&    99.83    &(-0.50)&    99.91    &(-0.25)\\
S      &   99.73   & (-0.28)&   100.04    &( 0.04)&    99.71    &(-0.31)&    99.95    &(-0.05)\\
H2S    &   99.80   & (-0.30)&    99.98    &(-0.03)&    99.75    &(-0.39)&    99.92    &(-0.12)\\
NO2    &   99.91   & (-0.45)&    99.98    &(-0.10)&    99.86    &(-0.69)&    99.92    &(-0.39)\\
O2     &  100.13   & ( 0.43)&   100.07    &( 0.24)&   100.08    &( 0.27)&   100.02    &( 0.06)\\
NO     &  100.02   & ( 0.07)&   100.03    &( 0.10)&    99.93    &(-0.21)&    99.98    &(-0.07)\\
N      &  100.17   & ( 0.13)&   100.17    &( 0.13)&   100.13    &( 0.10)&   100.00    &( 0.00)\\
H2O2   &   99.90   & (-0.37)&    99.97    &(-0.11)&    99.84    &(-0.57)&    99.91    &(-0.32)\\
N2     &   99.92   & (-0.21)&    99.99    &(-0.03)&    99.85    &(-0.39)&    99.93    &(-0.18)\\
SO3    &   99.81   & (-1.22)&    99.95    &(-0.35)&    99.81    &(-1.21)&    99.91    &(-0.59)\\
SO2    &   99.81   & (-0.89)&    99.95    &(-0.23)&    99.80    &(-0.99)&    99.91    &(-0.45)\\
HOCl   &   99.83   & (-0.57)&    99.96    &(-0.12)&    99.81    &(-0.64)&    99.91    &(-0.30)\\
HNO3   &   99.82   & (-1.17)&    99.94    &(-0.38)&    99.80    &(-1.37)&    99.88    &(-0.78)\\
N2H4   &   99.91   & (-0.27)&    99.96    &(-0.11)&    99.87    &(-0.41)&    99.91    &(-0.28)\\
NH2    &  100.00   & ( 0.00)&    99.99    &(-0.02)&    99.97    &(-0.05)&    99.96    &(-0.05)\\
Si2H6  &   99.82   & (-0.38)&    99.97    &(-0.07)&    99.80    &(-0.43)&    99.91    &(-0.20)\\
SiH3   &   99.81   & (-0.19)&    99.93    &(-0.07)&    99.73    &(-0.27)&    99.85    &(-0.15)\\
SH     &   99.81   & (-0.25)&   100.07    &( 0.08)&    99.70    &(-0.39)&    99.94    &(-0.08)\\
CH3SH  &   99.85   & (-0.42)&    99.96    &(-0.11)&    99.81    &(-0.53)&    99.91    &(-0.25)\\
CS     &   99.84   & (-0.36)&    99.98    &(-0.04)&    99.77    &(-0.55)&    99.93    &(-0.17)\\
CH3OH  &   99.93   & (-0.23)&    99.97    &(-0.10)&    99.89    &(-0.35)&    99.92    &(-0.27)\\
HCHO   &   99.92   & (-0.24)&    99.98    &(-0.07)&    99.86    &(-0.40)&    99.92    &(-0.23)\\
NH     &  100.11   & ( 0.12)&   100.10    &( 0.11)&    99.94    &(-0.07)&    99.93    &(-0.08)\\
SiH4   &   99.85   & (-0.17)&    99.97    &(-0.03)&    99.83    &(-0.21)&    99.91    &(-0.10)\\
Si     &   99.71   & (-0.15)&    99.68    &(-0.17)&    99.92    &(-0.04)&    99.98    &(-0.01)\\
C2H4   &   99.94   & (-0.15)&    99.97    &(-0.07)&    99.89    &(-0.29)&    99.92    &(-0.21)\\
CH3CH  &   99.90   & (-0.39)&    99.96    &(-0.15)&    99.86    &(-0.58)&    99.91    &(-0.38)\\
CS2    &   99.77   & (-0.85)&    99.96    &(-0.14)&    99.72    &(-1.03)&    99.90    &(-0.36)\\
NH3    &   99.96   & (-0.08)&    99.97    &(-0.05)&    99.90    &(-0.17)&    99.93    &(-0.13)\\
C      &  100.23   & ( 0.14)&   100.23    &( 0.14)&    99.95    &(-0.03)&    99.98    &(-0.01)\\
S2     &   99.65   & (-0.90)&    99.95    &(-0.14)&    99.60    &(-1.06)&    99.90    &(-0.26)\\
HCN    &   99.94   & (-0.15)&    99.99    &(-0.03)&    99.88    &(-0.31)&    99.94    &(-0.16)\\
C2H2   &   99.94   & (-0.13)&    99.99    &(-0.03)&    99.90    &(-0.23)&    99.94    &(-0.14)\\

RMSD:  &&            (0.44)    &&          (0.15)    &&          (0.53)    &&      (0.25)\\
Max:   &&           (-1.22)    &&          (0.47)    &&         (-1.37)    &&     (-0.78)\\

{} & {} & {} & {} & {} & {} & {} & {} & {}\\
\end{tabular}
\end{ruledtabular}
\label{Tab:table1}
\end{table}
Table ~\ref{Tab:table1} shows the percent recovery of DLPNO-CCSD(T)$_{\overline{\text{F12}}}$ correlation energies of all the molecular species 
present in the test set (excluding the hydrogen atom) with respect to the canonical CCSD(T)$_{\overline{\text{F12}}}$ values, where 
the numbers in the parentheses denote the errors in kcal/mol. For the aVTZ basis, the RMSD and max errors for {\tt TightPNO} 
and {\tt VeryTightPNO} settings are (0.44,-1.22) and (0.15,0.47) kcal/mol while for the aVQZ basis, the errors come out to be 
(0.53,-1.37) and (0.25,-0.78) kcal/mol respectively. 
\begin{figure}
\includegraphics[width=15cm]{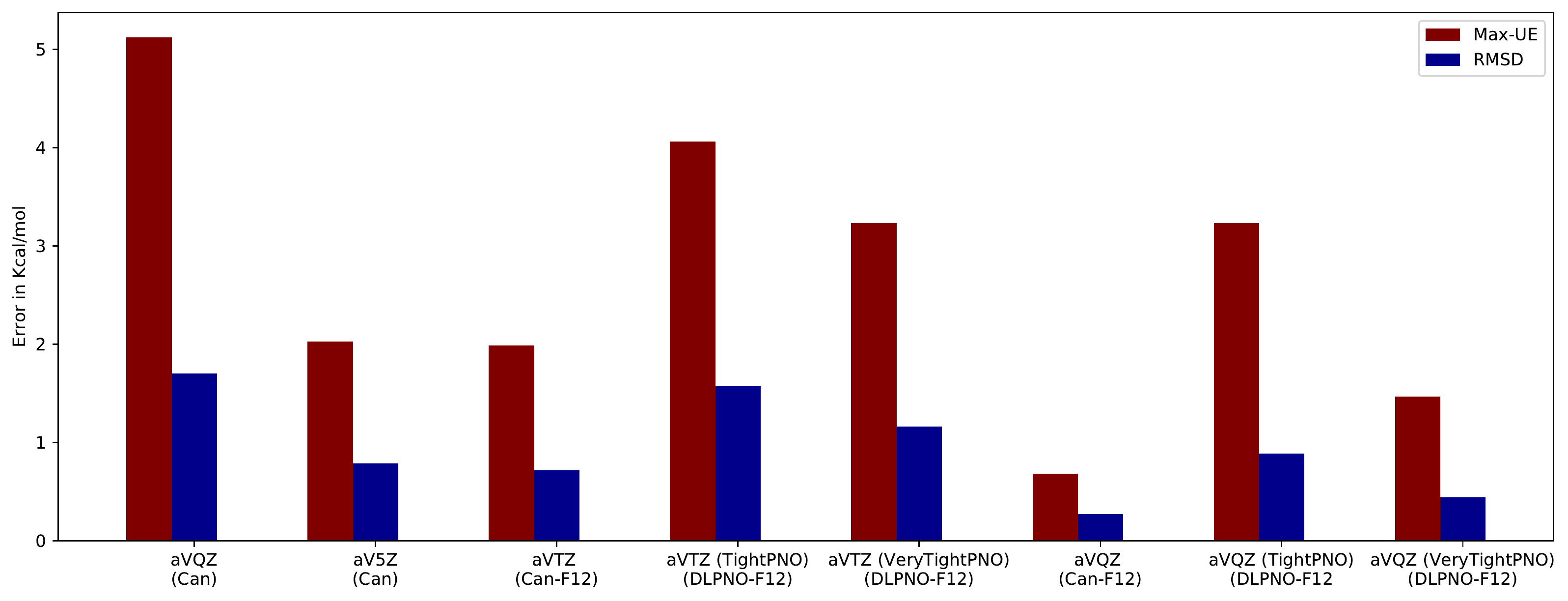}
\centering
\captionsetup{justification=raggedright,singlelinecheck=false}
\caption{{\footnotesize Max unsigned error and RMSD in kcal/mol in the reaction energies of the test set for the 
canonical CCSD(T) (Can), CCSD(T)$_{\overline{\text{F12}}}$ (Can-F12) and DLPNO-CCSD(T)$_{\overline{\text{F12}}}$ (DLPNO-F12) methods with respect to the 
CBS values.}}
\label{fig:cbs_compare_f12_werner}
\end{figure}
Fig.~\ref{fig:cbs_compare_f12_werner} illustrates the max unsigned error and RMSD of REs of 
the test set for the canonical CCSD(T) (Can), CCSD(T)$_{\overline{\text{F12}}}$ (Can-F12) and CCSD(T)$_{\overline{\text{F12}}}$ (DLPNO-F12) 
methods with respect to the CBS values. The basis set limit was estimated by using the Helgaker's two point extrapolation formula\cite{Helgaker:1997}
using aVQZ and aV5Z canonical CCSD(T) correlation energies. The HF energy and the direct singles term ($F^i_a t^a_i$)
was taken from the larger basis set. The Can-F12 method with aVTZ basis set is seen to perform 
slightly better than its conventional counterpart with aV5Z basis set: (0.72,1.99) vs (0.79,2.03) kcal/mol. However, these errors
are far from satisfactory and reflect the highly sensitive nature of these REs to the basis set size.
For the same test set, Werner reported improved results of (0.53,1.34) kcal/mol with his iterative CCSD(T)-F12b/aVTZ approach
which usually gives very similar results to our CCSD(T)$_{\overline{\text{F12}}}$ method for closed-shell systems\cite{Knizia:2009}.
The DLPNO-F12 method on the other hand resulted in quite large errors for the 
same basis set: (1.58,4.06) for {\tt TightPNO} and (1.16,3.23) kcal/mol for {\tt VeryTightPNO} settings. Thus, the aVTZ basis set is unable to provide 
the desired accuracy with both canonical and DLPNO-F12 methods. Finally, the Can-F12 method with aVQZ basis set reduces the errors further 
to (0.27,0.68) kcal/mol while the DLPNO approximations yield errors of (0.89,3.23) and (0.44,1.47) kcal/mol with {\tt TightPNO} and {\tt VeryTightPNO} 
settings respectively. Werner on the other hand obtained errors of (0.20,0.50) kcal/mol with the CCSD(T)-F12b/aVQZ method\cite{Knizia:2009}.
\\ This relatively poor performance of the DLPNO approach can be attributed to the PNOs being a suboptimal 
representation for capturing the F12 correlation effects. This is specially true for the PNOs of SOMO-DOMO pairs, 
for example, for the C atom in the aVTZ basis 10 PNOs contribute as much as 12\% of the F12 correlation energy for 
the SOMO-DOMO pairs in spite of having zero pair-density eigenvalue and having almost negligible contribution to the DLPNO-CCSD(T) correlation energy
(This will be elaborated in more detail in our future work). Furthermore, the errors in REs are further amplified by the lack of intrinsic error cancellations 
as the DLPNO-CCSD(T)$_{\overline{\text{F12}}}$ correlation energies of closed shell and most of the open shell species approach towards the canonical limit from 
above (<100\%) and below respectively (see Table ~\ref{Tab:table1}). This primarily happens due to the overestimation of the F12 correlation energy 
for the SOMO-DOMO pairs in the truncated PNO basis. The errors in REs are thus the largest for the smaller systems where the SOMO-DOMO F12 correlation 
energies constitute a significant portion of the total F12 correction and hence a bigger basis set like aVQZ with {\tt VeryTightPNO} truncation parameters 
is required for a favorable comparison with the CBS values for such systems. We also plan to employ open-shell geminal spanning orbitals (GSOs), earlier proposed by some 
of us\cite{Pavosevic:2014} within the DLPNO framework as an alternative to PNOs in order to cut down the computational costs associated with the {\tt VeryTightPNO}
truncation settings.

\subsection {Computational scaling}
In this section, we analyze the scaling behavior of both DLPNO-CCSD(T) and DLPNO-CCSD(T)$_{\overline{\text{F12}}}$ methods for open-shell systems,
while employing the less expensive {\tt T0} variant of (T) (due to the compute time constraints) although similar trends are expected to hold for 
the rigorous ({\tt T1}) variant as well. We have chosen the perturbative triples method (instead of just CCSD) as well in this analysis as the 
F12 correction is often used in conjunction with the CCSD(T) procedure.
\begin{figure}
\includegraphics[width=8cm]{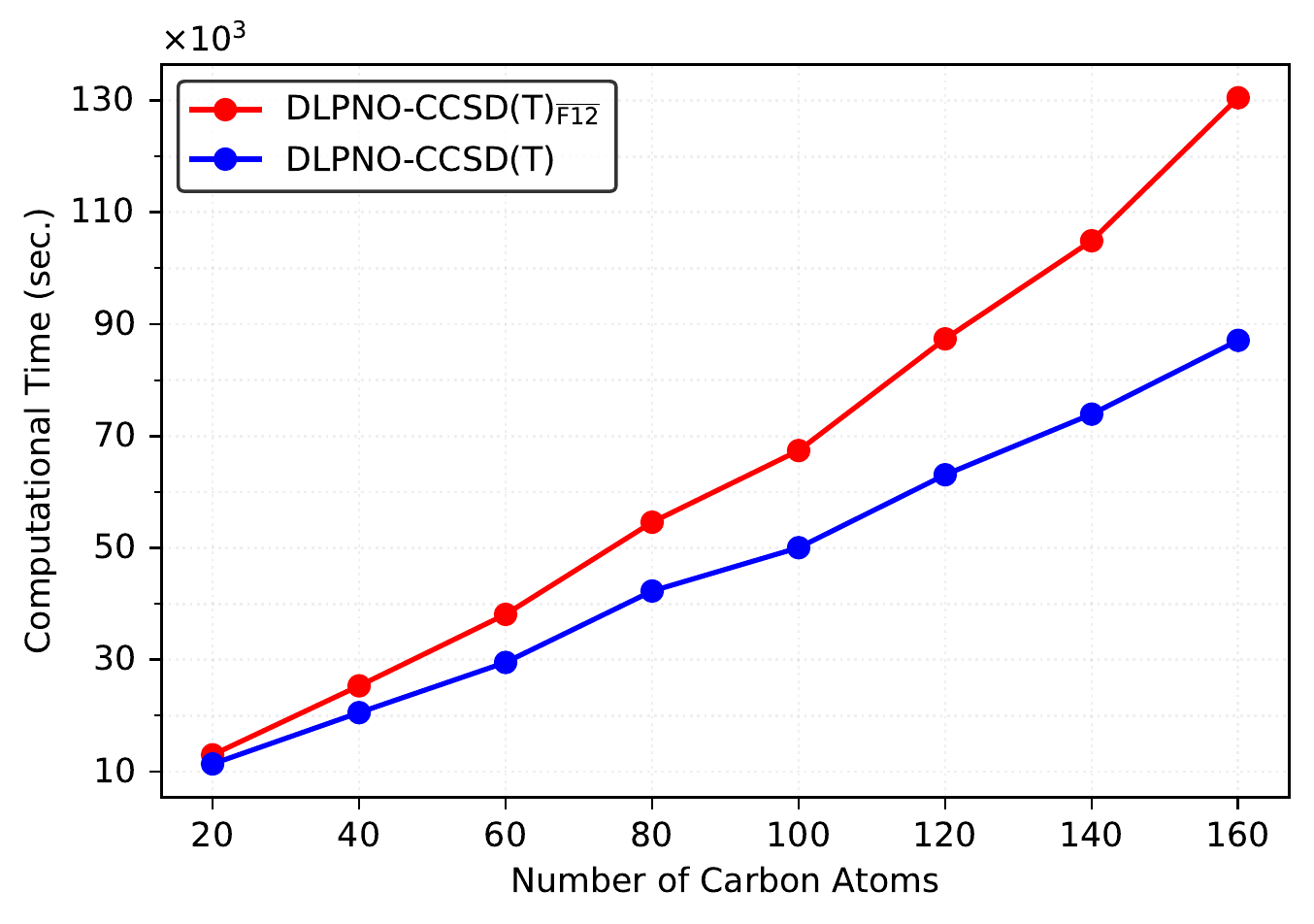}
\centering
\captionsetup{justification=raggedright,singlelinecheck=false}
\caption{{\footnotesize Wall clock time in seconds for UHF-DLPNO-CCSD(T) and UHF-DLPNO-CCSD(T)$_{\overline{\text{F12}}}$ correlation energy calculations
with {\tt TightPNO} settings on quasilinear n-alkane chains $C_{\text{n}}H_{\text{2n+2}}$, in triplet state from $n$=20 to $n$=160. 
Following basis set quartet were used: \{def2-TZVP, def2-TZVP/RI, cc-pVDZ-F12-OptRI, def2/J\} \{OBS, DFBS, CABS, CDFBS\}
and the COSX procedure for evaluating the exchange operator was employed. All these calculations utilized 4 CPU cores (4 MPI processes) and a total of 512 GB memory.}}
\label{fig:scaling}
\end{figure}
Fig.~\ref{fig:scaling} shows the wall clock time in seconds for both UHF-DLPNO-CCSD(T) and UHF-DLPNO-CCSD(T)$_{\overline{\text{F12}}}$
calculations on quasilinear n-alkane chains $C_{\text{n}}H_{\text{2n+2}}$ in triplet state, 
from $n$=120 to $n$=160. These calculations utilized 4 CPU cores (4 MPI processes) and a total of 512 GB memory. The effective exponents
for the UHF-DLPNO-CCSD(T) and UHF-DLPNO-CCSD(T)$_{\overline{\text{F12}}}$ calculations from Fig.~\ref{fig:scaling} came out to be \textbf{1.06} and \textbf{1.23} respectively.
The deviation from linear scaling for the UHF-DLPNO-CCSD(T)$_{\overline{\text{F12}}}$ method can be attributed to almost quadratic scaling 1-external RI integral generation
in the crude guess step\cite{Saitow:2017bo} and the cubically scaling density fitted evaluation of the Coulomb operator ($\hat{J}$) in OBS (O) and CABS (C) spaces: 
$\hat{J}_{OO}$, $\hat{J}_{CO}$ and $\hat{J}_{CC}$. Furthermore, some of the I/O steps in the pair-specific PNO integral generation procedure in F12
(which number as $2\times N_{\text{Pairs}}$) doesn't scale linearly as we go to more than 100 atoms. We hope to address these terms in the near future.
In \ce{C160H322}, the time spent in F12 procedure is mostly divided into 3 parts: 3-index RI integral transformation to PAO basis (39\%), the Fock matrix 
construction (27\%) and the evaluation of intermediates in PNO basis(31\%) (refer to Table\ref{tab:table2}) and the entire F12 procedure took about 33\% of the total
computational time. We also looked at the scaling behavior of our method for a series of triplet 
[(${\text{C}_{\text{4}}\text{SH}_{\text{3}})-(\text{CH}_{\text{2}})_{N}-(\text{C}_{\text{4}}\text{SH}_{\text{3}})}^{2+}]$ diradical molecules (with $N$=0..50) 
previously used to study computational scaling
with the system size\cite{Guo:2016DLPNONEVPT2} using identical
settings as above (refer to the Supporting Information for the 
corresponding plot).
We also compared both RHF and UHF variants of DLPNO-CCSD(T)$_{\overline{\text{F12}}}$ for n-alkanes as above in the singlet state. 
\begin{figure}[ht] 
  \begin{subfigure}[b]{0.5\linewidth}
    \centering
    \includegraphics[width=0.85\linewidth]{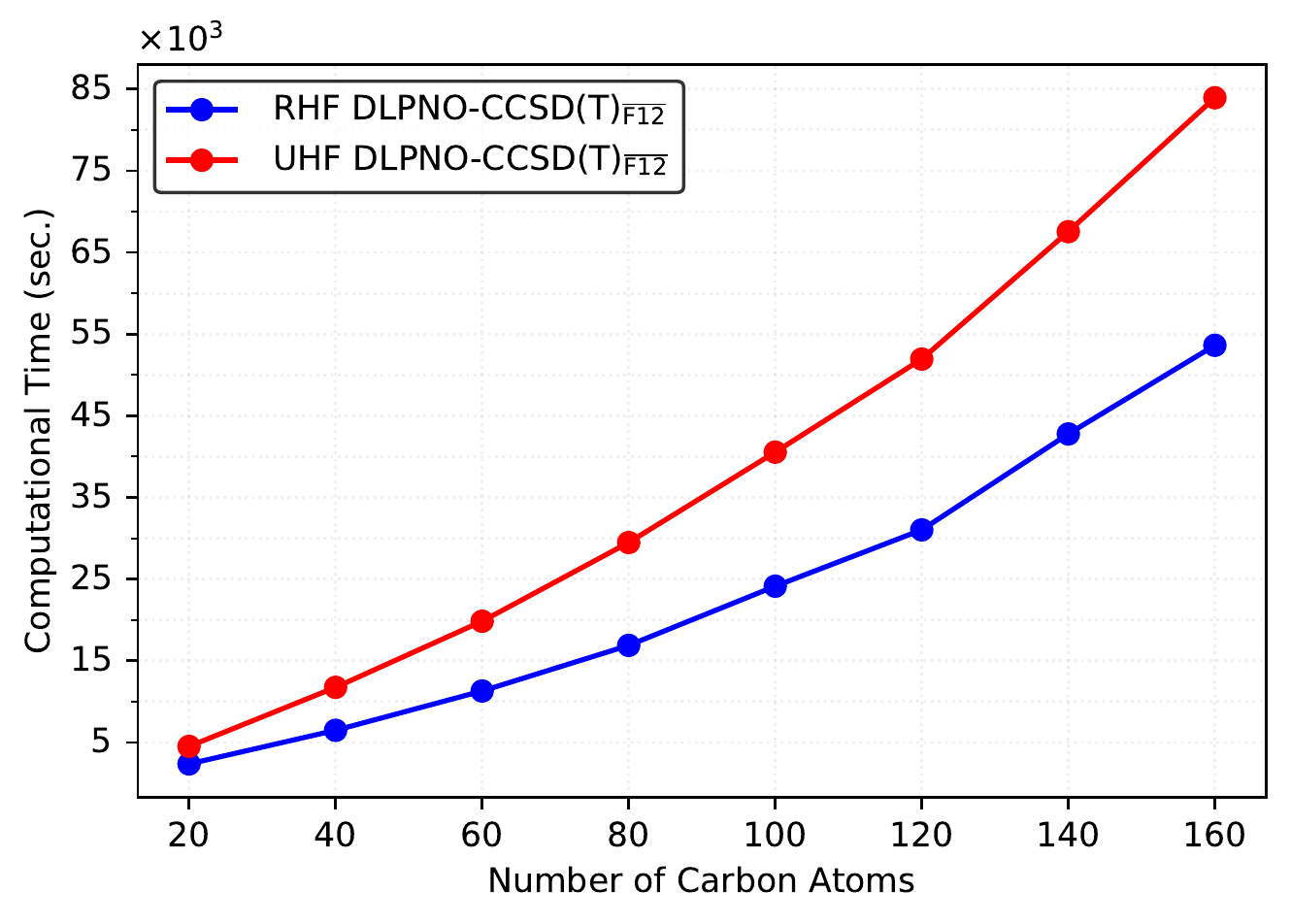} 
    \caption{RHF/UHF DLPNO-CCSD(T)$_{\overline{\text{F12}}}$} 
    \label{fig4:a} 
  \end{subfigure}
  \begin{subfigure}[b]{0.5\linewidth}
    \centering
    \includegraphics[width=0.85\linewidth]{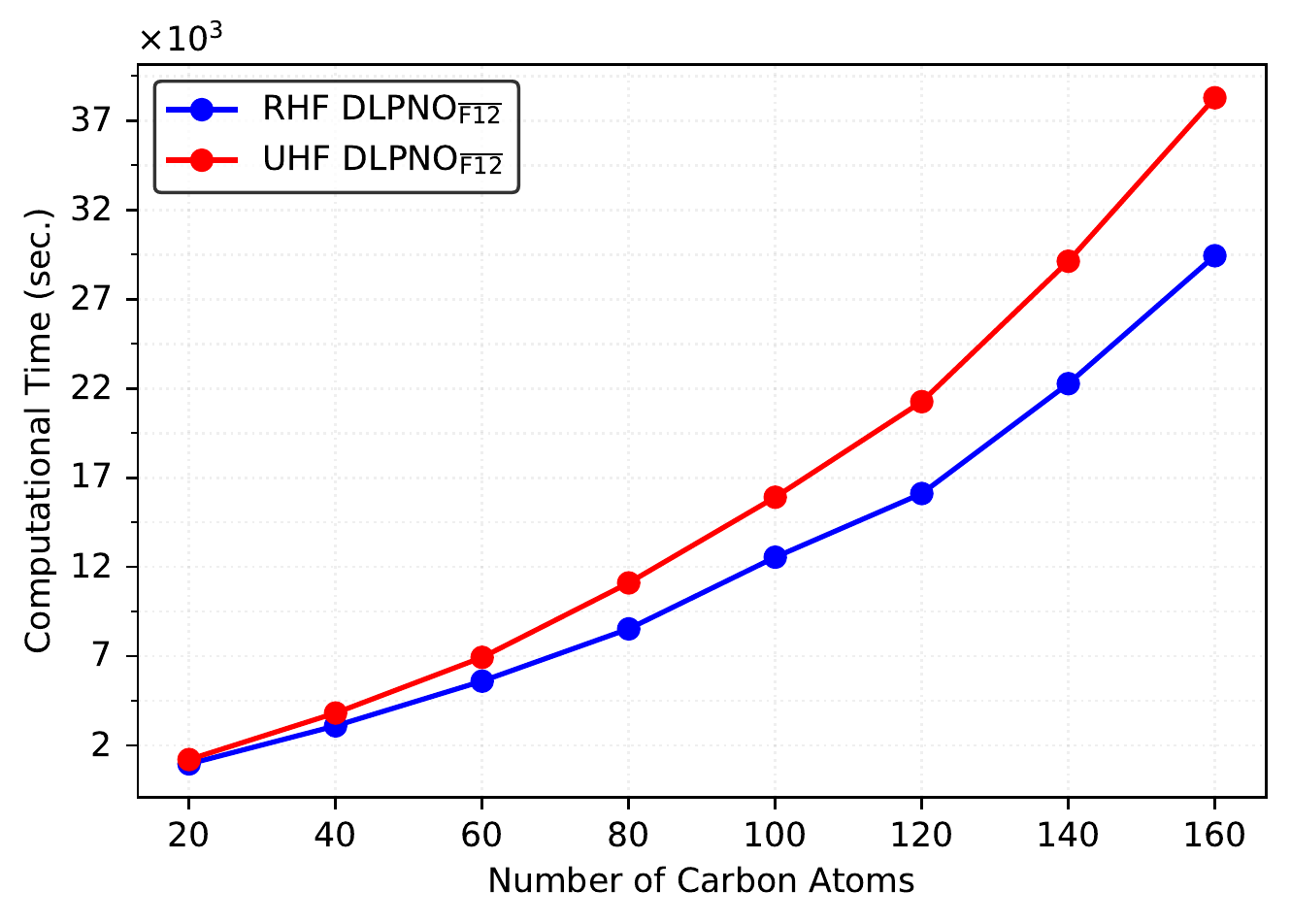} 
    \caption{RHF/UHF DLPNO-F12} 
    \label{fig4:b} 
  \end{subfigure} 
  \captionsetup{justification=raggedright,singlelinecheck=false}
\caption{{\footnotesize a) Total and b) F12-only computation time (in seconds) for closed- and open-shell DLPNO-CCSD(T)$_{\overline{\text{F12}}}$ implementations with {\tt TightPNO} settings on quasilinear 
n-alkane chains \ce{C$_n$H$_{2n+2}$}, in singlet state from $n$=20 to 
$n$=160. Following basis set quartet were used: \{def2-TZVP, def2-TZVP/RI, cc-pVDZ-F12-OptRI, def2/J\} \{OBS, DFBS, CABS, CDFBS\} 
and the COSX procedure for evaluating the exchange operator was employed. All these calculations utilized 4 CPU cores (4 MPI processes) and a total of 512 GB memory.}}
\label{fig4} 
\end{figure}
The correlation energies obtained from both closed- and open-shell DLPNO-CCSD(T)$_{\overline{\text{F12}}}$ methods match up to 5 decimal points, signifying balanced treatment of open- and closed-shell systems by
our approach. From Fig.~\ref{fig4:a} and Fig.~\ref{fig4:b}, it can be
seen that the UHF code is only fractionally more expensive than the RHF code. Specifically, for $\text{C}_{160}\text{H}_{322}$ alkane, the UHF-DLPNO-CCSD(T)
method is only 1.5 times its closed shell counterpart even though the CCSD iterations in the open-shell could be 3-4 times more expensive 
due to the appearance of more pairs ($\alpha-\alpha, \alpha-\beta, \beta-\beta$). Furthermore, if we look at only the F12 part of calculations
(Fig.~\ref{fig4:b}), the difference is even smaller and the UHF code is only expensive 1.3 times for the same alkane, helped by the fact that 
one doesn't need to solve 3-sets of CC amplitude-like equations in the F12 part. For a small number of SOMOs, both integral transformations and evaluation of 
F12 intermediates is only marginally expensive (compared to closed-shell) in the open-shell case. Of course, additional (spin) evaluations of the exchange 
operator in different spaces are required. A favorable comparison with the closed-shell code makes this a successful extension of the RHF-DLPNO-CCSD(T)$_{\overline{\text{F12}}}$ 
formalism to open-shell systems.

\subsection{Medium and large molecules}
In this section, we analyze the timings of sample UHF-DLPNO-CCSD(T)$_{\overline{\text{F12}}}$ calculations on representative
medium-sized and large molecular species which are currently beyond the reach of the corresponding canonical formalisms. Their 
optimized geometries were taken from the SI of Ref\cite{Saitow:2017bo}. Dioxygenate, Vitamine E succinate, Cochineal Carmine and Hexanitrohexaazaisowurtzitane 
(CL20) constitutes our set of medium sized molecules. Both the transition-state ({\em ts}) and educt models of 
of the dioxygenase enzyme active site\cite{Neese:2012bio} are considered.
Model compounds for the active center of the [\ce{NiFe}] hydrogenase enzyme\cite{Neese:2013bio} (triplet state) and bicarbonate in Photosystem II\cite{Athina:2013bio} were chosen
as test-cases for assessing the performance of the UHF-DLPNO-CCSD(T)$_{\overline{\text{F12}}}$ method for larger systems. For all the transition metal complexes, QROs obtained from open-shell 
density functional calculations with CAM-B3LYP functional\cite{Handy:2004} 
were used as reference orbitals. Furthermore, the RIJCOSX approximation (using def2/J as
CDFBS) was used to evaluate the Fock matrix. Since optimized CABS basis 
sets are not available for transition metals like \ce{Fe} and \ce{Ni}, the {\em uncontracted} def2-TZVPP (bicarbonate) and def2-QZVPP 
(dioxygenate, [\ce{NiFe}] hydrogenase) basis sets were used as the 
CABS. While, medium-sized molecular calculations used the {\tt TightPNO} truncation settings, {\tt NormalPNO} settings were used for the bigger calculations. Also, the 
semi-canonical non-iterative ({\tt T0}) variant of the (T) procedure was employed. All calculations utilized 4 CPU cores (4 MPI processes) and 512GB memory. 
Table~\ref{tab:table2} shows the breakdown of wall time (in seconds) for different (prominent) computational steps involved in the UHF-DLPNO-CCSD(T)$_{\overline{\text{F12}}}$ procedure 
excluding the preceding reference calculations. Timings of \ce{C160H322} molecule has also been added for comparison purposes. It can be seen that the time spent in 
the F12 procedure for the medium sized molecules is only a tiny fraction of the total execution time and ranges from 3.8\% in educt dioxygenate to 12.7\% in 
Vitamin E succinate. For the \ce{C160H322} molecule, however, the F12 contribution is a bit higher at 33.2\% because of the comparatively faster (T) 
calculation. For the larger systems, only DLPNO-CCSD$_{\overline{\text{F12}}}$ calculations were carried out and the time spent in the F12 part come out to be 39.8\% and 40.2\% for 
Bicarbonate and [\ce{NiFe}] hydrogenase molecules respectively. Obviously, this will be lowered significantly if
the (T) correction was included (especially, the iterative (T1) variety),
as it would normally be for practical applications that call for the high-precision provided by the explicitly-correlated coupled-cluster.
The F12 correction 
for the Bicarbonate system took around 26 hours, and is the most extensive
F12 calculation reported in this work, including more than 500 atoms. Thus clearly very large systems can be already treated by the presented open-shell DLPNO-CCSD(T)$_{\overline{\text{F12}}}$ approach.
However there is additional untapped potential for optimization in our implementation, besides obvious ones (like large-scale parallelization).
Inside the F12 part of the calculation, a significant portion of time is spent in the Fock matrix formation, integral transformations in the PAO 
space, calculation of pair-specific CABS space, pair-specific OBS/OBS and CABS/OBS integral generations in PNO basis and the evaluation of the $\bar{V}$ intermediate.
As mentioned before, the I/O in the OBS/OBS and CABS/OBS integral generation routine can still be brought down significantly. Currently, the pair-extended integrals
in the PAO space are read from disk for every occupied orbital $i$ and $j$ appearing in the F12 pair list, resulting in a 2 $\times N_{\text{Pairs}}$ I/O operations, followed by 
integral sieving to extract only the $i$ and $j$ specific integrals which are then finally fused to construct the integrals associated with the $ij$ pair. This could be changed into a 
having a loop over occupied orbitals $i$, and then generating all the OBS/OBS and CABS/OBS integrals for all the $j$ orbitals connected to the orbital $i$. This way, 
the number of I/O steps would be reduced to the number of occupied orbitals, which in principle, should lower down the prefactor of the PNO integral generation 
part of the code significantly. Also, similar to the works of Tew at al.\cite{Schmitz:2014}, the generation of the CABS space could be made cheaper by making use of 
a more compact representation for the CABS orbitals like OSVs than just the regular AOs. 

These optimizations will be pursued further in future work.

\clearpage
\newpage
\pagestyle{empty}
\begin{turnpage}
\begin{table}
\tiny
\captionsetup{font=scriptsize,justification=raggedright,singlelinecheck=false}
\caption{\label{tab:table2}
Computational timings (in seconds) for the various steps of the DLPNO-CCSD(T)$_{\overline{\text{F12}}}$ procedure along with reference, correlation and F12 correction energies for representative medium-sized
and large open-shell systems. Reference calculations on all molecules (except \ce{C160H322}) were done using UKS/CAM-B3LYP. RIJCOSX approximation was used in all these calculations (using def2/J basis set) and each calculation utilized 4 CPU cores (4 MPI processes) and a total of 512 GB memory.}
\begin{ruledtabular}
\begin{tabular}{lcccccccc}
{} &  Dioxygenate (ts)\footnotemark[1]&  Dioxygenate (educt)\footnotemark[1] &    Vitamin E\footnotemark[2] &  Cochineal\footnotemark[2] 
& CL20\footnotemark[2] & C$_{160}$H$_{322}$\footnotemark[4] & Bicarbonate\footnotemark[5] & [\ce{NiFe}] hydrogenase\footnotemark[6]\\
\midrule
Total execution time:                                &       17318&       19344&        60724&       130252&      234992&        130492&        239032&        130505\\
Time in CCSD procedure:                              &        8594&       10996&        22006&        40985&       82225&         68081&        143888&         78000\\
Time in (T) procedure:                               &        7979&        7611&        31002&        78122&      142888&         19031&           -  &           -  \\
		                                     &            &            &             &             &            &              &              &              \\
Total time and \% in F12 procedure:                  & 744 (4.3\%)& 736 (3.8\%)&7716 (12.7\%)&11144 (8.6\%)&9878 (4.2\%)&43380 (33.2\%)&95144 (39.8\%)&52504 (40.2\%)\\
		   	                             &            &            &             &             &            &              &              &              \\
\hspace{2mm} Fock matrix formation                   &          50&          50&          504&          406&         245&         11776&         12513&          4026\\
\hspace{2mm} Integral transformations (PAO)          &         141&         141&         2596&         2208&        1122&         16897&         67788&         26531\\
\hspace{2mm} Pair-specific 1e matrix                 &          24&          24&          505&          875&         706&           872&          1614&          2505\\
\hspace{2mm} Local fit metric generation             &          35&          33&          120&          270&         309&            68&            91&           310\\
\hspace{2mm} Calculation of CABS                     &          32&          32&          668&         1144&         818&          1951&          3097&          3295\\
\hspace{2mm} OBS/OBS integral generation             &          30&          30&          218&          380&         372&          1570&           748&          1441\\
\hspace{2mm} CABS/OBS integral generation            &          84&          84&          938&         1512&        1386&          3346&          5237&          4754\\
\hspace{2mm} $V_{ij}, fX_{ij}, B_{ij}$ Intermediates &          16&          17&          156&          303&         332&           263&           790&           748\\
\hspace{2mm} $\bar{V}_{ij}$ Intermediate   	     &         292&         287&         1808&         3606&        4202&          1631&          1950&          7932\\
						     &            &            &             &             &            &              &              &              \\
Reference energy:                                    &-1940.059567&-1940.064296& -1276.512050& -1818.281941&-1781.258625&  -6248.368368& -15181.158996&  -7843.018084\\
F12 correction to reference energy:                  &   -0.037876&   -0.037977&    -0.076530&    -0.102759&   -0.111546&     -0.357284&    -11.205403&     -0.163318\\
CCSD Correlation energy:                             &   -3.068416&   -3.066244&    -5.480703&    -6.370557&   -6.035779&    -28.781440&    -46.386367&    -17.248012\\
F12 correction to CCSD correction energy:            &   -0.314036&   -0.314318&    -0.610031&    -0.838685&   -0.783082&     -2.326865&    -11.221550&     -1.321897\\
(T) correlation energy:                              &   -0.135943&   -0.133316&    -0.243998&    -0.309254&   -0.340798&     -1.125222&           -  &        -     \\
Total energy:                                        &-1943.615838&-1943.616152& -1282.923311& -1825.903196&-1788.529831&  -6280.959179& -15249.972316&  -7861.751311\\
						     &            &            &             &             &            &              &              &              \\
no. of SOMOs:      				     &           6&           6&            1&            1&           1&             2&             4&             2\\
no. of atoms:      				     &          22&          22&           80&           55&          36&           482&           565&           180\\
						     &            &            &             &             &            &              &              &              \\
size of AO basis:   				     &         545&         545&         1371&         1230&         954&          6892&          5420&          4007\\
size of DF basis:   				     &        1352&        1352&         3359&         2980&        2298&         16990&         17743&          9594\\ 
size of CABS basis:\footnotemark[7] 		     &        1351&        1351&         4872&         4335&        3348&         27426&         17702&          9810\\
\end{tabular}
\end{ruledtabular}
\renewcommand{\footnotesize}{\scriptsize}
\footnotetext[1]{\{def2-TZVPP, def2-TZVPP/RI, {\em uncontracted} def2-QZVPP\}, {\tt TightPNO}}
\footnotetext[2]{\{cc-pVDZ-F12, aug-cc-pVDZ/RI, cc-pVDZ-F12-OptRI\}, {\tt TightPNO}}
\footnotetext[4]{\{def2-TZVP, def2-TZVP/RI, cc-pVDZ-F12-OptRI\}, {\tt TightPNO}}
\footnotetext[5]{\{def2-SVP, def2-SVP/RI, {\em uncontracted} def2-TZVPP\}, {\tt NormalPNO}}
\footnotetext[6]{\{def2-TZVPP, def2-TZVPP/RI, {\em uncontracted} def2-QZVPP\}, {\tt NormalPNO}}
\footnotetext[7]{CABS space includes AOs as well in CABS+ approach}
\end{table}
\end{turnpage}
\clearpage
\newpage
\pagestyle{plain}
\section{Conclusion}\label{sec:Summary}
We have developed a robust explicitly correlated DLPNO-CCSD(T)$_{\overline{\text{F12}}}$ method 
for high-spin open-shell species whose costs and storage scale nearly linearly with the system size.
For representative applications we found that the time spent inside the F12 part of the calculations is only a fraction of the total computation time. We were able to carry out F12 calculations on the model 
compound of bicarbonate present in the PSII enzyme, which involved over 5000 basis-functions and more than 500 atoms, which to our knowledge, 
is the most extensive open-shell F12 calculation to this date.

The approach allows to compute CCSD(T) energies that approach the complete basis set limit very closely.
For example, when using only a aug-cc-pVTZ basis set and the {\tt TightPNO} 
truncation settings (default for F12)
the DLPNO-CCSD(T)$_{\overline{\text{F12}}}$ heats of formation of 50 largest molecules out of the ANL database of 348 core combustion species are within
0.4 kcal/mol of the reference extrapolated CBS CCSD(T) values.
The deviation is further lowered to $\approx 0.3$ kcal/mol with extrapolation of DLPNO-CCSD(T)$_{\overline{\text{F12}}}$
results with aug-cc-pVDZ and aug-cc-pVTZ basis sets. We also examined the reaction energies of 50 open-shell reactions 
with large reaction energies and strong basis set dependence where we obtained a RMSD of $\sim$0.4 kcal/mol 
with the aug-cc-pVQ(+d)Z basis set. However, {\tt VeryTightPNO} truncation settings were employed for these calculations
resulting in a significant increase in the associated computational costs. The use of open-shell GSOs\cite{Pavosevic:2014} will be explored in our 
next work as an alternative to PNOs as they have been shown to be much more compact for describing F12 intermediates.
Thus, the DLPNO-CCSD(T)$_{\overline{\text{F12}}}$ method is an attractive alternative to purely extrapolation based canonical
and reduced-scaling CCSD(T) formalisms. The F12/D approximation\cite{Pavosevic:2016bc} (used in all the F12 calculations reported in the paper), which avoids the need 
for explicit evaluation of the exchange operator in the CABS-CABS space and is hence computationally cheaper, is recommended along with the {\tt TightPNO} 
truncation settings. A production-quality implementation of the DLPNO-CCSD(T)$_{\overline{\text{F12}}}$ method has been implemented in the 4.2 version of the ORCA 
quantum chemistry package, making it suitable for a widespread use in the chemistry community. Further improvements in the DLPNO F12 code like the introduction of GSOs\cite{Pavosevic:2014}, compact representation 
of the CABS space, and reduction of I/O steps in the evaluation of intermediates are expected in the near future.
\section*{Supplementary Material}
See supplementary material for a) individual heats of formation of the 50 largest molecules of the ANL database 
and b) the scaling plot of the DLPNO-CCSD(T)$_{\overline{\text{F12}}}$ method for a series of triplet 
[(${\text{C}_{\text{4}}\text{SH}_{\text{3}})-(\text{CH}_{\text{2}})_{N}-(\text{C}_{\text{4}}\text{SH}_{\text{3}})}^{2+}]$ diradical molecules.

\begin{acknowledgments}
This work was supported by the U.S. National Science Foundation (awards 1550456 and 1800348) and by the Exascale Computing Project (17-SC-20-SC), a collaborative effort of
the U.S. Department of Energy Office of Science and the National Nuclear Security Administration.
The authors acknowledge Virginia Tech's Advanced Research Computing center (ARC) for providing computational resources and technical support that made this work possible.
\end{acknowledgments}

~\\
~\\
{\bf Data Availability Statement}
~\\
The data that supports the findings of this study are available within the article and its supplementary material while the (remaining) 
raw data are available from the corresponding author upon request.

\nocite{*}
\bibliography{main}

\end{document}


\title{Supporting Information: Near-linear scaling explicitly correlated coupled cluster method based on an open-shell domain-based local pair natural orbitals}

\author{Ashutosh Kumar} 
\affiliation{Department of Chemistry, Virginia Tech, Blacksburg, Virginia 24061, United States}

\author{Frank Neese}
\affiliation{Max-Planck-Institut f{\"u}r Chemische Energiekonversion, Stiftstr. 34-36, D-45470 M{\"u}lheim an der Ruhr, Germany}

\author{Edward F. Valeev}
 \email{efv@vt.edu}
\affiliation{Department of Chemistry, Virginia Tech, Blacksburg, Virginia 24061, United States}

\date{\today} 
\maketitle
\begin{figure}
\includegraphics[width=8cm]{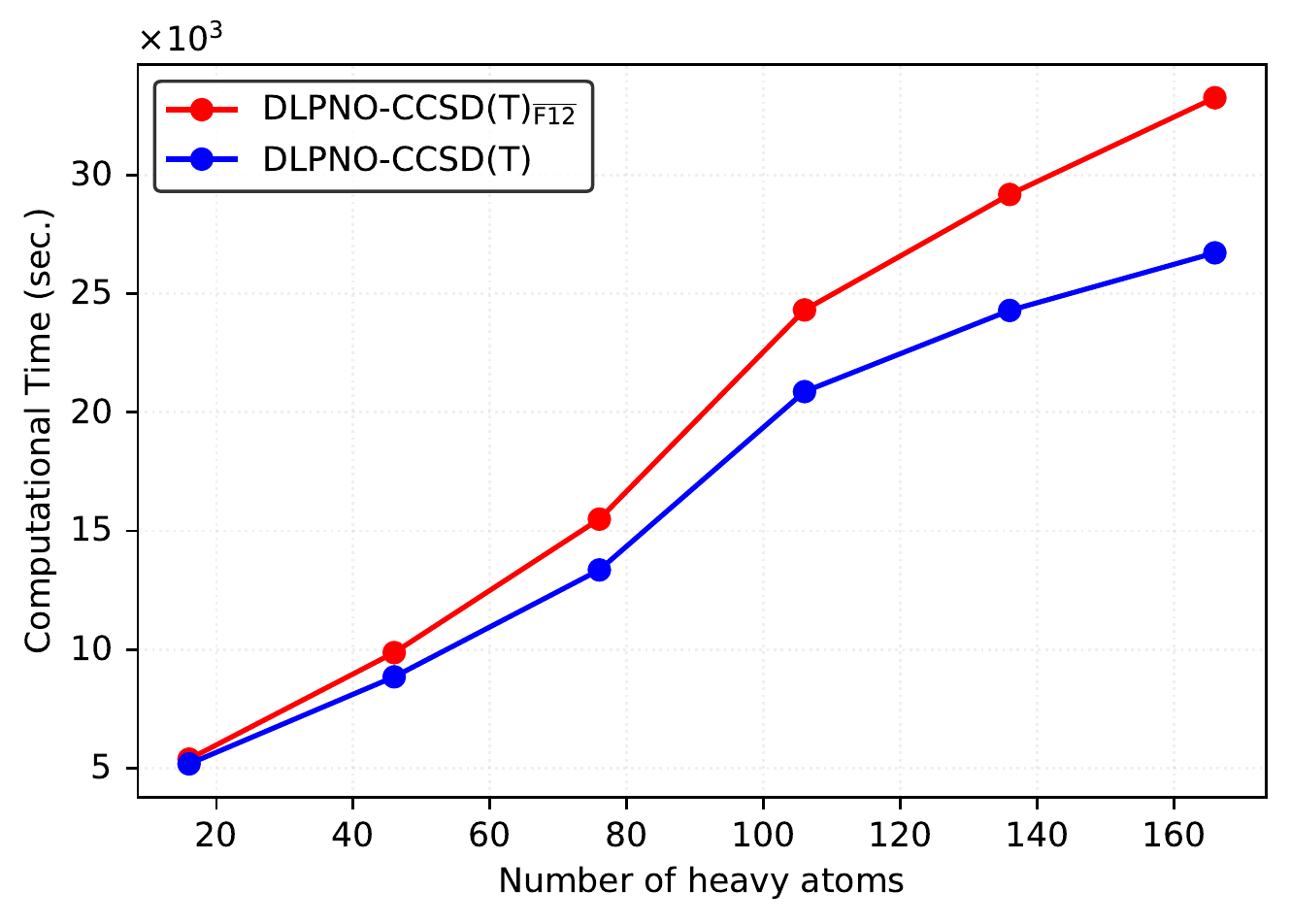}
\centering
\captionsetup{justification=raggedright,singlelinecheck=false}
\caption{{\footnotesize Wall clock time in seconds for UHF-DLPNO-CCSD(T) and UHF-DLPNO-CCSD(T)$_{\overline{\text{F12}}}$ correlation energy calculations
with {\tt TightPNO} settings for a series of triplet [(${\text{C}_{\text{4}}\text{SH}_{\text{3}})-(\text{CH}_{\text{2}})_{N}-(\text{C}_{\text{4}}\text{SH}_{\text{3}})}^{2+}]$ diradical molecules from $N$=0 to $N$=50.
Following basis set quartet were used: \{def2-TZVP, def2-TZVP/RI, cc-pVDZ-F12-OptRI, def2/J\} \{OBS, DFBS, CABS, CDFBS\}
and the COSX procedure for evaluating the exchange operator was employed. All these calculations utilized 4 CPU cores (4 MPI processes) and a total of 512 GB memory.}}
\label{fig:scaling_2}
\end{figure}

\begin{table} 
\tiny
\captionsetup{font=scriptsize,justification=raggedright,singlelinecheck=false}
\caption{Comparison of heats of formation in kcal/mol calculated using the {\tt T0} and {\tt T1} approximations of the DLPNO-CCSD(T)$_{\overline{\text{F12}}}$ method with {\tt TightPNO} settings for 
cc-pVDZ-F12 and cc-pVTZ-F12 basis set with the CBS extrapolated canonical CCSD(T) values. Errors with respect to the CBS results are shown in the parenthesis.}
\begin{ruledtabular}
\begin{tabular}{lccccccccc}
 {} &  \multicolumn{4}{c}{cc-pVDZ-F12\footnotemark[1]}  & \multicolumn{4}{c}{cc-pVTZ-F12\footnotemark[2]}  & CBS\footnotemark[3]\\
\cmidrule(r){2-5}\cmidrule(l){6-9} 
 {} &  \multicolumn{2}{c}{$T_0$}  & \multicolumn{2}{c}{$T_1$}  & \multicolumn{2}{c}{$T_0$} & \multicolumn{2}{c}{$T_1$}  & \\
\bottomrule
{} & {} & {} & {} & {} & {} & {} & {} & {} & {} \\
CH3CHOO      &  35.15    &( 2.13)&    34.24    &(-1.22)&    35.17    &(-2.15)&    34.29    &(-1.27)&    33.02\\ 
CH2CHOOH     &  14.32    &(-0.59)&    13.82    &(-0.09)&    14.70    &(-0.97)&    14.20    &(-0.47)&    13.73\\
CH3CH2OO     &  14.13    &(-0.02)&    13.87    &( 0.24)&    14.55    &(-0.44)&    14.31    &(-0.20)&    14.11\\
OHCH2CH2O    & -20.14    &( 0.11)&   -20.27    &( 0.24)&   -19.87    &(-0.16)&   -19.99    &(-0.04)&   -20.03\\
OHCH2OCH2    & -24.15    &(-0.61)&   -24.34    &(-0.42)&   -24.18    &(-0.58)&   -24.36    &(-0.40)&   -24.76\\
CH3OCH2O     & -15.89    &( 0.03)&   -16.05    &( 0.19)&   -15.48    &(-0.38)&   -15.63    &(-0.23)&   -15.86\\
OHCHCH2OH    & -31.38    &(-0.70)&   -31.56    &(-0.52)&   -31.58    &(-0.50)&   -31.75    &(-0.33)&   -32.08\\
(CH3)3C      &  34.95    &(-0.35)&    34.85    &(-0.25)&    34.73    &(-0.13)&    34.63    &(-0.03)&    34.60\\
CH3CH2CHCH3  &  38.49    &(-0.35)&    38.40    &(-0.26)&    38.31    &(-0.17)&    38.22    &(-0.08)&    38.14\\
CH3CH2CHOH   &   2.52    &(-0.53)&     2.38    &(-0.39)&     2.32    &(-0.33)&     2.19    &(-0.20)&     1.99\\
CH2CCCCH     & 190.24    &( 0.47)&   189.35    &( 1.36)&   191.18    &(-0.47)&   190.29    &( 0.42)&   190.71\\
CHCCHCCH     & 192.04    &( 0.45)&   191.22    &( 1.27)&   192.93    &(-0.44)&   192.12    &( 0.37)&   192.49\\
CH3OOO       &  26.56    &(-0.02)&    25.31    &( 1.23)&    28.18    &(-1.64)&    26.97    &(-0.43)&    26.54\\
OCH2OOH      &  -2.32    &( 0.37)&    -2.56    &( 0.61)&    -1.50    &(-0.45)&    -1.72    &(-0.23)&    -1.95\\
OHCH2OO      & -21.80    &( 0.11)&   -22.09    &( 0.40)&   -21.18    &(-0.51)&   -21.45    &(-0.24)&   -21.69\\
C(OH)3       & -79.77    &(-0.64)&   -79.97    &(-0.44)&   -79.81    &(-0.60)&   -80.01    &(-0.40)&   -80.41\\
(CH3)2CHCH2  &  40.04    &(-0.39)&    39.93    &(-0.28)&    39.82    &(-0.17)&    39.71    &(-0.06)&    39.65\\
CH3CH(OH)CH2 &   6.26    &(-0.50)&     6.13    &(-0.37)&     6.09    &(-0.33)&     5.96    &(-0.20)&     5.76\\
CH3OCHOH     & -25.29    &(-0.59)&   -25.48    &(-0.40)&   -25.27    &(-0.61)&   -25.46    &(-0.42)&   -25.88\\
CH2CH2CH2OH  &   8.77    &(-0.45)&     8.65    &(-0.33)&     8.63    &(-0.31)&     8.51    &(-0.19)&     8.32\\
CH3N(O)OH    &  13.77    &(-0.45)&    13.36    &(-0.04)&    14.16    &(-0.84)&    13.77    &(-0.45)&    13.32\\
CH3CHCH2OH   &   6.77    &(-0.41)&     6.65    &(-0.29)&     6.66    &(-0.30)&     6.55    &(-0.19)&     6.36\\
CHCCCO       & 144.04    &( 0.46)&   142.85    &( 1.65)&   145.45    &(-0.95)&   144.28    &( 0.22)&   144.50\\
CH2CH2OOH    &  33.13    &(-0.30)&    32.90    &(-0.07)&    33.37    &(-0.54)&    33.16    &(-0.33)&    32.83\\
CH2CH2OCH3   &  19.95    &(-0.49)&    19.81    &(-0.35)&    19.88    &(-0.42)&    19.74    &(-0.28)&    19.46\\
CH3C(OH)CH3  &  -2.38    &(-0.52)&    -2.53    &(-0.37)&    -2.60    &(-0.30)&    -2.74    &(-0.16)&    -2.90\\
CH3CH2OCH2   &  13.33    &(-0.73)&    13.17    &(-0.57)&    13.09    &(-0.49)&    12.93    &(-0.33)&    12.60\\
CH3CH(O)CH3  &   9.81    &( 0.08)&     9.70    &( 0.19)&     9.96    &(-0.07)&     9.85    &( 0.04)&     9.89\\
CH3CH2CH2O   &  12.48    &( 0.17)&    12.37    &( 0.28)&    12.69    &(-0.04)&    12.60    &( 0.05)&    12.65\\
CH3CHOCH3    &  11.59    &(-0.62)&    11.43    &(-0.46)&    11.43    &(-0.46)&    11.27    &(-0.30)&    10.97\\
CHCCCOH      & 126.92    &(-0.48)&   125.86    &( 0.58)&   127.54    &(-1.10)&   126.48    &(-0.04)&   126.44\\
CH2CCCO      & 103.73    &(-0.72)&   101.92    &( 1.09)&   104.99    &(-1.98)&   103.19    &(-0.18)&   103.01\\
CH(OH)3      &-129.63    &(-0.98)&  -129.85    &(-0.76)&  -129.88    &(-0.73)&  -130.10    &(-0.51)&  -130.61\\
OHCH2OOH     & -60.21    &(-0.80)&   -60.52    &(-0.49)&   -60.13    &(-0.88)&   -60.43    &(-0.58)&   -61.01\\
OHOOOH       &   5.95    &(-0.48)&     5.35    &( 0.12)&     7.01    &(-1.54)&     6.42    &(-0.95)&     5.47\\
3c-CHCHCHCHC & 172.90    &(-0.40)&   172.19    &( 0.31)&   173.53    &(-1.03)&   172.83    &(-0.33)&   172.50\\
CH2CCCCH2    & 158.54    &(-0.65)&   157.06    &( 0.83)&   159.63    &(-1.74)&   158.13    &(-0.24)&   157.89\\
CHCCH2CCH    & 158.87    &(-0.20)&   158.03    &( 0.64)&   159.56    &(-0.89)&   158.71    &(-0.04)&   158.67\\
CH3CCCCH     & 149.13    &(-0.50)&   148.12    &( 0.51)&   149.69    &(-1.06)&   148.67    &(-0.04)&   148.63\\
OCCHCCH      &  98.64    &(-0.17)&    97.52    &( 0.95)&    99.71    &(-1.24)&    98.59    &(-0.12)&    98.47\\
CH3CH2OOH    & -21.64    &(-0.81)&   -21.92    &(-0.53)&   -21.68    &(-0.77)&   -21.95    &(-0.50)&   -22.45\\
CH3CH2CH2CH3 & -12.74    &(-0.83)&   -12.89    &(-0.68)&   -13.17    &(-0.40)&   -13.31    &(-0.26)&   -13.57\\
(CH3)2CHCH3  & -14.16    &(-0.86)&   -14.31    &(-0.71)&   -14.63    &(-0.39)&   -14.79    &(-0.23)&   -15.02\\
CH3OOCH3     & -12.07    &(-0.83)&   -12.37    &(-0.53)&   -11.98    &(-0.92)&   -12.29    &(-0.61)&   -12.90\\
OCCCO        &  25.57    &(-0.27)&    24.06    &( 1.24)&    26.78    &(-1.48)&    25.29    &( 0.01)&    25.30\\
CH3CH2OCH3   & -35.56    &(-0.92)&   -35.74    &(-0.74)&   -35.84    &(-0.64)&   -36.03    &(-0.45)&   -36.48\\
CH2CCHCCH    & 155.00    &(-0.31)&   154.01    &( 0.68)&   155.81    &(-1.12)&   154.80    &(-0.11)&   154.69\\
c-CHCHCHCHNH &  68.51    &(-1.41)&    67.70    &(-0.60)&    68.57    &(-1.47)&    67.75    &(-0.65)&    67.10\\
CHCCCCCH     & 233.28    &(-0.59)&   231.67    &( 1.02)&   234.29    &(-1.60)&   232.66    &( 0.03)&   232.69\\
CHCCCCN      & 212.62    &( 0.45)&   210.96    &( 2.11)&   214.45    &(-1.38)&   212.78    &( 0.29)&   213.07\\
{} & {} & {} & {} & {} & {} & {} & {} & {}& {}\\
Max          & &  (-2.13) & &  ( 2.11) & &   (-2.15) & &   (-1.27) & \\
RMSD         & &  ( 0.63) & &  ( 0.73) & &   ( 0.92) & &   ( 0.38) & \\
\end{tabular}
\end{ruledtabular}
\label{Tab:table1}
\renewcommand{\footnotesize}{\scriptsize}
\footnotetext[1]{\{cc-pVDZ-F12, cc-pVDZ-F12-OptRI, aug-cc-pVDZ/RI\}}
\footnotetext[2]{\{cc-pVTZ-F12, cc-pVTZ-F12-OptRI, aug-cc-pVTZ/RI\}}
\footnotetext[3]{\{a$'$5Z/a$'$QZ extrapolation\}}
\end{table}
\begin{table} 
\tiny
\captionsetup{font=scriptsize,justification=raggedright,singlelinecheck=false}
\caption{Comparison of heats of formation in kcal/mol calculated using the {\tt T0} and {\tt T1} approximations of the DLPNO-CCSD(T)$_{\overline{\text{F12}}}$ method 
with {\tt TightPNO} settings for aug-cc-pVDZ and aug-cc-pVTZ basis with the CBS extrapolated canonical CCSD(T) values. Errors with respect to the CBS results are shown in the parenthesis.}
\begin{ruledtabular}
\begin{tabular}{lccccccccc}
 {} &  \multicolumn{4}{c}{aug-cc-pVDZ\footnotemark[1]}  & \multicolumn{4}{c}{aug-cc-pVTZ\footnotemark[2]}  & CBS\footnotemark[3]\\
\cmidrule(r){2-5}\cmidrule(l){6-9} 
 {} &  \multicolumn{2}{c}{$T_0$}  & \multicolumn{2}{c}{$T_1$}  & \multicolumn{2}{c}{$T_0$} & \multicolumn{2}{c}{$T_1$}  & \\
\bottomrule
{} & {} & {} & {} & {} & {} & {} & {} & {} & {} \\
CH3CHOO      &   35.17    &(-2.15)&    34.26    &(-1.24)&    34.83    &(-1.81)&    33.94    &(-0.92)&    33.02\\ 
CH2CHOOH     &   14.18    &(-0.45)&    13.68    &( 0.05)&    14.48    &(-0.75)&    13.98    &(-0.25)&    13.73\\
CH3CH2OO     &   13.33    &( 0.78)&    13.08    &( 1.03)&    14.02    &( 0.09)&    13.77    &( 0.34)&    14.11\\
OHCH2CH2O    &  -20.70    &( 0.67)&   -20.83    &( 0.80)&   -20.13    &( 0.10)&   -20.24    &( 0.21)&   -20.03\\
OHCH2OCH2    &  -24.76    &( 0.00)&   -24.95    &( 0.19)&   -24.40    &(-0.36)&   -24.59    &(-0.17)&   -24.76\\
CH3OCH2O     &  -16.59    &( 0.73)&   -16.75    &( 0.89)&   -15.87    &( 0.01)&   -16.02    &( 0.16)&   -15.86\\
OHCHCH2OH    &  -31.98    &(-0.10)&   -32.15    &( 0.07)&   -31.71    &(-0.37)&   -31.87    &(-0.21)&   -32.08\\
(CH3)3C      &   34.61    &(-0.01)&    34.49    &( 0.11)&    34.89    &(-0.29)&    34.79    &(-0.19)&    34.60\\
CH3CH2CHCH3  &   38.17    &(-0.03)&    38.06    &( 0.08)&    38.44    &(-0.30)&    38.35    &(-0.21)&    38.14\\
CH3CH2CHOH   &    2.10    &(-0.11)&     1.96    &( 0.03)&     2.32    &(-0.33)&     2.18    &(-0.19)&     1.99\\
CH2CCCCH     &  193.39    &(-2.68)&   192.47    &(-1.76)&   191.94    &(-1.23)&   191.04    &(-0.33)&   190.71\\
CHCCHCCH     &  195.10    &(-2.61)&   194.26    &(-1.77)&   193.72    &(-1.23)&   192.90    &(-0.41)&   192.49\\
CH3OOO       &   25.28    &( 1.26)&    24.00    &( 2.54)&    27.29    &(-0.75)&    26.07    &( 0.47)&    26.54\\
OCH2OOH      &   -3.04    &( 1.09)&    -3.28    &( 1.33)&    -2.07    &( 0.12)&    -2.29    &( 0.34)&    -1.95\\
OHCH2OO      &  -22.67    &( 0.98)&   -22.96    &( 1.27)&   -21.83    &( 0.14)&   -22.11    &( 0.42)&   -21.69\\
C(OH)3       &  -80.47    &( 0.06)&   -80.67    &( 0.26)&   -80.10    &(-0.31)&   -80.29    &(-0.12)&   -80.41\\
(CH3)2CHCH2  &   39.55    &( 0.10)&    39.43    &( 0.22)&    39.95    &(-0.30)&    39.84    &(-0.19)&    39.65\\
CH3CH(OH)CH2 &    5.72    &( 0.04)&     5.58    &( 0.18)&     6.10    &(-0.34)&     5.97    &(-0.21)&     5.76\\
CH3OCHOH     &  -25.84    &(-0.04)&   -26.03    &( 0.15)&   -25.53    &(-0.35)&   -25.72    &(-0.16)&   -25.88\\
CH2CH2CH2OH  &    8.27    &( 0.05)&     8.15    &( 0.17)&     8.64    &(-0.32)&     8.52    &(-0.20)&     8.32\\
CH3N(O)OH    &   13.17    &( 0.15)&    12.77    &( 0.55)&    13.63    &(-0.31)&    13.24    &( 0.08)&    13.32\\
CH3CHCH2OH   &    6.40    &(-0.04)&     6.27    &( 0.09)&     6.69    &(-0.33)&     6.57    &(-0.21)&     6.36\\
CHCCCO       &  146.68    &(-2.18)&   145.47    &(-0.97)&   146.04    &(-1.54)&   144.86    &(-0.36)&   144.50\\
CH2CH2OOH    &   32.53    &( 0.30)&    32.31    &( 0.52)&    33.16    &(-0.33)&    32.94    &(-0.11)&    32.83\\
CH2CH2OCH3   &   19.43    &( 0.03)&    19.28    &( 0.18)&    19.80    &(-0.34)&    19.66    &(-0.20)&    19.46\\
CH3C(OH)CH3  &   -2.74    &(-0.16)&    -2.89    &(-0.01)&    -2.59    &(-0.31)&    -2.73    &(-0.17)&    -2.90\\
CH3CH2OCH2   &   12.68    &(-0.08)&    12.51    &( 0.09)&    12.96    &(-0.36)&    12.80    &(-0.20)&    12.60\\
CH3CH(O)CH3  &    9.29    &( 0.60)&     9.17    &( 0.72)&     9.88    &( 0.01)&     9.77    &( 0.12)&     9.89\\
CH3CH2CH2O   &   11.98    &( 0.67)&    11.87    &( 0.78)&    12.57    &( 0.08)&    12.47    &( 0.18)&    12.65\\
CH3CHOCH3    &   11.28    &(-0.31)&    11.12    &(-0.15)&    11.36    &(-0.39)&    11.20    &(-0.23)&    10.97\\
CHCCCOH      &  129.46    &(-3.02)&   128.39    &(-1.95)&   128.14    &(-1.70)&   127.07    &(-0.63)&   126.44\\
CH2CCCO      &  105.78    &(-2.77)&   103.95    &(-0.94)&   105.33    &(-2.32)&   103.52    &(-0.51)&   103.01\\
CH(OH)3      & -130.66    &( 0.05)&  -130.87    &( 0.26)&  -130.27    &(-0.34)&  -130.49    &(-0.12)&  -130.61\\
OHCH2OOH     &  -61.28    &( 0.27)&   -61.58    &( 0.57)&   -60.65    &(-0.36)&   -60.96    &(-0.05)&   -61.01\\
OHOOOH       &    4.77    &( 0.70)&     4.18    &( 1.29)&     6.14    &(-0.67)&     5.55    &(-0.08)&     5.47\\
3c-CHCHCHCHC &  174.11    &(-1.61)&   173.39    &(-0.89)&   174.05    &(-1.55)&   173.34    &(-0.84)&   172.50\\
CH2CCCCH2    &  160.98    &(-3.09)&   159.48    &(-1.59)&   159.99    &(-2.10)&   158.49    &(-0.60)&   157.89\\
CHCCH2CCH    &  161.47    &(-2.80)&   160.61    &(-1.94)&   160.21    &(-1.54)&   159.35    &(-0.68)&   158.67\\
CH3CCCCH     &  151.90    &(-3.27)&   150.87    &(-2.24)&   150.29    &(-1.66)&   149.26    &(-0.63)&   148.63\\
OCCHCCH      &  100.73    &(-2.26)&    99.60    &(-1.13)&   100.12    &(-1.65)&    99.00    &(-0.53)&    98.47\\
CH3CH2OOH    &  -22.52    &( 0.07)&   -22.79    &( 0.34)&   -22.08    &(-0.37)&   -22.35    &(-0.10)&   -22.45\\
CH3CH2CH2CH3 &  -13.46    &(-0.11)&   -13.62    &( 0.05)&   -13.23    &(-0.34)&   -13.37    &(-0.20)&   -13.57\\
(CH3)2CHCH3  &  -14.99    &(-0.03)&   -15.16    &( 0.14)&   -14.67    &(-0.35)&   -14.83    &(-0.19)&   -15.02\\
CH3OOCH3     &  -12.92    &( 0.02)&   -13.22    &( 0.32)&   -12.44    &(-0.46)&   -12.74    &(-0.16)&   -12.90\\
OCCCO        &   27.06    &(-1.76)&    25.52    &(-0.22)&    27.16    &(-1.86)&    25.66    &(-0.36)&    25.30\\
CH3CH2OCH3   &  -36.21    &(-0.27)&   -36.39    &(-0.09)&   -36.08    &(-0.40)&   -36.27    &(-0.21)&   -36.48\\
CH2CCHCCH    &  157.40    &(-2.71)&   156.39    &(-1.70)&   156.38    &(-1.69)&   155.37    &(-0.68)&   154.69\\
c-CHCHCHCHNH &   68.98    &(-1.88)&    68.17    &(-1.07)&    68.73    &(-1.63)&    67.92    &(-0.82)&    67.10\\
CHCCCCCH     &  237.50    &(-4.81)&   235.86    &(-3.17)&   235.17    &(-2.48)&   233.53    &(-0.84)&   232.69\\
CHCCCCN      &  216.77    &(-3.70)&   215.10    &(-2.03)&   215.27    &(-2.20)&   213.59    &(-0.52)&   213.07\\
{} & {} & {} & {} & {} & {} & {} & {} & {}& {}\\
Max          & &  (-4.81) & &  (-3.17) & &   (-2.48) & &   (-0.92) & \\
RMSD         & &  ( 1.64) & &  ( 1.11) & &   ( 1.06) & &   ( 0.40) & \\
\end{tabular}
\end{ruledtabular}
\label{Tab:table2}
\renewcommand{\footnotesize}{\scriptsize}
\footnotetext[1]{\{aug-cc-pVDZ, cc-pVDZ-F12-OptRI, aug-cc-pVDZ/RI\}}
\footnotetext[2]{\{aug-cc-pVTZ, cc-pVDZ-F12-OptRI, aug-cc-pVTZ/RI\}}
\footnotetext[3]{\{a$'$5Z/a$'$QZ extrapolation\}}
\end{table}